From naculich@Bowdoin.EDU Thu Jan 20 13:53:58 2000
Date: Thu, 20 Jan 2000 13:53:56 -0500
From: naculich@Bowdoin.EDU
To: lozano@brandeis.edu

\documentstyle[11pt]{article}

\input{psfig}

\def\text{}
\renewcommand{\baselinestretch}{1.75} 

\newcommand{\al}{\alpha}

\newcommand{\th}{\theta}

\newcommand{\la}{\lambda}

\newcommand{\De}{\Delta}

\newcommand{\La}{\Lambda}

\def\ie{{\it i.e.}}

\def\quarS{{\bar S}}
\def\ieject{}
\def\arh{ r}
\def\thh{ {\textstyle {3\over 2}} }

\def\tshalf{ {\textstyle {1\over 2}} }
\def\ha{ {\textstyle {1\over 2}} }
\def\xe{2}
\def\xtw{}
\def\xf{}
\def\nha{\ha}
\def\xxt{}

\def\NSone{ {\bigcirc \!\!\!\!\! 1}\,\,}
\def\NStwo{ {\bigcirc \!\!\!\!\! 2}\,\,}
\def\NSthree{ {\bigcirc \!\!\!\!\! 3}\,\,}
\def\NSfour{ {\bigcirc \!\!\!\!\! 4}\,\,}

\newcommand{\be}{\begin{eqnarray}}
\newcommand{\ee}{\end{eqnarray}}

\newcommand{\tf}{\tilde{f}}

\newcommand{\pr}{\partial}

\newcommand{\np}{\newpage}
\newcommand{\hs}{\hspace}
\newcommand{\vs}{\vspace}
\newcommand{\nl}{\newline}
\newcommand{\nn}{\nonumber}

\newcommand{\RR}{{\rm I\kern-1.6pt {\rm R}}}

\newcommand{\ZZ}{{\rm Z}\kern-3.8pt {\rm Z} \kern2pt}

\begin{document}

\thispagestyle{empty}

\vs*{-25mm}
\begin{flushright}
BRX-TH-468\\[-.2in]
HUTP-99/A068\\[-.2in]
BOW-PH-116\\[-.2in]
\end{flushright}
\vspace{.3in}
\setcounter{footnote}{0}

\begin{center}
{\Large{\bf Elliptic models and M-theory}}
\renewcommand{\baselinestretch}{1}
\small
\normalsize
\vspace{.3in}

Isabel P. Ennes\footnote{
Research supported 
by the DOE under grant DE--FG02--92ER40706.}$^{,a}$, 
Carlos Lozano\footnotemark[1]$^{,a}$, Stephen G. Naculich\footnote{
Research supported in part by the National Science Foundation under grant 
no.~PHY94-07194 through the \\
\phantom{aaa}  ITP Scholars Program.}$^{,b}$, 
Howard J. Schnitzer\footnote{Permanent address.}
${}^{\!\!\!,\!\!\!}$
\footnote{Research supported in part
by the DOE under grant DE--FG02--92ER40706.\\
{\tt \phantom{aaa} naculich@bowdoin.edu; 
ennes,lozano,schnitzer@brandeis.edu}\\}$^{,a,c}$\\

\vspace{.2in}

$^{3,a}$Martin Fisher School of Physics\\
Brandeis University, Waltham, MA 02454

\vspace{.2in}

$^{b}$Department of Physics\\
Bowdoin College, Brunswick, ME 04011

\vspace{.2in}

$^{c}$Lyman Laboratory of Physics \\
Harvard University, Cambridge, MA 02138

\vspace{.3in}

{\bf{Abstract}} 
\end{center}
\renewcommand{\baselinestretch}{1.75}
\small
\normalsize
\begin{quotation}
\baselineskip14pt
\noindent  

We give a unified analysis of four-dimensional elliptic models
with ${\cal N}=2$ supersymmetry and a simple gauge group,
and their relation to M-theory. 
Explicit calculations of the Seiberg-Witten curves 
and the resulting one-instanton prepotential are presented.
The remarkable regularities that emerge are emphasized. 
In addition, we calculate the prepotential in the Coulomb phase 
of the (asymptotically-free) Sp$(2N)$ gauge theory 
with $N_f$ fundamental hypermultiplets of arbitrary mass.

\end{quotation}

\np 

\setcounter{page}{1}

\noindent{\bf 1. ~Introduction}
\renewcommand{\theequation}{1.\arabic{equation}}
\setcounter{equation}{0}

The study of ${\cal N}=2$ supersymmetric gauge theories using the
Seiberg-Witten (SW) approach~\cite{SeibergWitten}  
to the low energy effective action is now more than five years old. 
During this period the theory has undergone considerable development, 
from a variety of approaches. 
One of the intriguing aspects of SW theory is the connection to
integrable models, where M-theory \cite{Witten} 
provides one method of constructing
the spectral curves of elliptic models.
(Another technique is geometric engineering  \cite{geomengineering}.)
Even though this aspect of elliptic models has been extensively studied
\cite{DonagiWitten}--\cite{Gorsky}, 
there remain a number of open questions of some importance for these theories. 
In particular, except in certain special cases, the bridge between the
spectral curve of the elliptic model and the corresponding curve
obtained from an M-theory picture is still absent. 
This is one of the issues we consider in this paper, 
with considerable progress, but not
a complete resolution of all the issues.

One motivation for understanding the connection 
between the spectral curve and M-theory picture 
is to present the instanton expansion of
the prepotential ${\cal F}$ for the theory in question. 
As explained in our previous papers in this series
\cite{oneanti}--\cite{santiago}, 
this will provide tests of M-theory by means of comparison 
between our results for 
${\cal F}_{\rm instanton}$ 
with the analogous instanton prepotential
obtained from the microscopic Lagrangian\footnote{
Slater \cite{slater} has calculated 
${\cal F}_{\rm 1-inst}$ for ${\cal N} = 2$ SU$(N)$ gauge theory with 
one symmetric hypermultiplet and $N_f$ fundamental hypermultiplets
using the microscopic Lagrangian. 
His result is in agreement with the predictions of refs.~\cite{onesym,nonhyper}
obtained using the M-theory curve of ref.~\cite{LandsteinerLopezLowe}.
This provides the first independent check of the predictions 
obtained using hyperelliptic perturbation
theory \cite{oneanti}.} \cite{micro}--\cite{slater}.

The breakthrough of Seiberg and Witten \cite{SeibergWitten} 
was their formulation of the exact solution of 
\hbox{4-dimensional}
\hbox{${\cal N}=2$} supersymmetric gauge theories 
in terms of a low-energy (Wilsonian) effective action
accurate to two derivatives of the fields,
\be 
{\cal L}_{\rm eff}\,=\,\frac{1}{4\pi} {\rm
Im}\left(\int {\rm d}^4\th \frac{\pr {\cal F}(A)}{\pr A_i}\bar{A_i}+
\frac{1}{2}\int {\rm d}^2\th\frac{\pr^2 {\cal F}(A)}{\pr A_i\,\pr
A_j} W^{\al}_i\,W_{\al,j}\right)\,+\,{\rm higher~derivatives}, 
\label{four} 
\ee 
where $A^i$ are ${\cal N}=1$ chiral superfields
($i=1 \,\,\,{\rm to}\,\, \,{\rm rank}\,{\cal G}$), 
${\cal F}(A)$ is the holomorphic prepotential, 
and $W^i$ is the gauge field strength.
The holomorphic prepotential can be expressed in terms of a
perturbative piece and an infinite series of instanton contributions 
\be 
{\cal F}(A)\,=\, {\cal F}_{\rm classical}(A)\, +\,{\cal F}_{\rm
1-loop}(A)\,+\,\sum _{d=1}^{\infty} L^{2d} {\cal
F}_{\rm d-inst}(A), 
\label{six} 
\ee 
where $L^2 = \Lambda^{I(G)-\sum_{R}I(R)}$ 
with $\Lambda$ the quantum scale (Wilson cutoff),  
$I(G)$ the Dynkin index of the adjoint representation,
and $I(R)$ the Dynkin index of a matter hypermultiplet in representation $R$. 
The one-loop contribution is given by perturbation theory
\be 
 {\cal F}_{\rm 1-loop}(a) &=&\frac{i}{4 \pi}\sum_{\al
\in \De_+} (a\cdot \al)^2\,{\rm log} \left(\frac{a\cdot
\al}{\La}\right)^2\nn \\ 
& &-\frac{i}{8 \pi} \sum_{j} \sum_{w \in R_j}
(a\cdot w+m_j)^2\, {\rm log}\left(\frac{a\cdot
w+m_j}{\La}\right)^2, 
\label{seven} 
\ee 
where $\alpha$ ranges over
the positive roots $\De_+$ of $\cal G$, 
$w$ runs over the weight vectors for a hypermultiplet with mass $m_j$
in the representation $R_j$, 
and $a_i$ parametrizes the Cartan subalgebra of ${\cal G}$. 
For models with zero beta function, 
the instanton expansion is in powers of $q$ rather than $\Lambda$,
where \hbox{$q=e^{2\pi i\tau}$} 
with $\tau$ the coupling constant of the theory.

In order to compute the prepotential (\ref{six}) 
using the Seiberg-Witten approach (for a recent review, see
\cite{DHPREV}), 
one requires:

\noindent (1) A suitable Riemann surface or algebraic curve, 
appropriate to the gauge group and matter content of the theory,
dependent on gauge invariant moduli $u_i$ 
(equivalently on the order parameters $a_i$)
and the masses of the hypermultiplets.

\noindent (2) A preferred meromorphic 1-form $\lambda$, the SW differential.

\noindent (3) A canonical basis of homology cycles $(A_k, B_k)$ on the surface.

\noindent  These data allow the computation of period integrals 
\be 
2\pi i a_k=\oint_{A_k}\la, \hs{15mm} 2\pi ia_{D,k}=\oint_{B_k}\la,
\label{eight} 
\ee 
from which one may compute ${\cal F}(a)$ by integrating 
$a_{D,k}\,=\,{{\pr {\cal F}{(a)}}\over {\pr a_k}}$.

In this paper we will discuss the SW theory for all simple classical groups
${\cal G}$, with matter hypermultiplets in the asymptotically free
Coulomb phase, or in the Coulomb phase with zero beta function. 
The discussion will be comprehensive in the sense that 
we will consider all generic cases 
({\ie} of arbitrary rank ${\cal G}$) 
for such models. 
The SW curves for these models fall into three classes: 

\noindent (a) hyperelliptic curves \cite{Mtheory, Everybody},  

\noindent (b) cubic (non-hyperelliptic) curves \cite{LandsteinerLopezLowe}
\footnote{The curve for SU($N$) + one antisymmetric representation 
was recently derived from an integrable model \cite{krichephong}.}, and 

\noindent (c) curves of infinite order. 

Our focus in this paper will be primarily on the last class of curves
which correspond to decompactified elliptic models.
The M-theory pictures for elliptic models for theories 
with vanishing beta function were given by Uranga \cite{Uranga};
those with simple classical groups are:

\noindent (1) SU($N$) with two antisymmetric hypermultiplets and four
fundamental hypermultiplets,

\noindent (2) SU($N$) with an antisymmetric and a symmetric hypermultiplet,

\noindent (3) SU($N$) with an adjoint hypermultiplet,

\noindent (4) SO($N$) with an adjoint hypermultiplet,

\noindent (5) Sp($2N$) with an adjoint hypermultiplet, and

\noindent (6) Sp($2N$) with an antisymmetric hypermultiplet and four
fundamental hypermultiplets.

We will explicitly write down the curves (leading-order terms only)
for these models,
and the resulting  one-instanton prepotential. 
By sending the masses of some of the fundamental hypermultiplets to 
infinity, we recover the curves for some additional models
in the Coulomb phase which also possess curves of infinite order.

A number of methods exist for extracting the instanton expansion 
from hyperelliptic curves, 
with the method of asymptotic expansion  
\cite{DHokerKricheverPhong1}--\cite{DHokerPhong} being the 
most useful  for our purposes. 
In refs.  \cite{oneanti}--\cite{twoanti},
we have extended these ideas to cases (b) and (c), 
developing methods for obtaining the instanton expansion for 
non-hyperelliptic SW curves, of finite or infinite order. 
In this body of work, 
the order parameter $a_k$ emerges as the natural variable 
for describing the instanton expansion, 
rather than the gauge invariant moduli.
See ref.~\cite{santiago} for a review and more details.

In section~2, we discuss Sp$(2N)$ gauge theory
with $N_f$ fundamental hypermultiplets of arbitrary masses,
resolving some issues that were left open by previous 
work \cite{DHokerKricheverPhong3}. 
In section~3, 
we assemble the results for the one-instanton prepotential 
for models with different groups and matter content,
observing a remarkable empirical regularity among the different cases.
In section~4,
we summarize the M-theory pictures for the decompactified elliptic models,
from which we obtain the leading-order terms (defined in sec.~4)
of the coefficient functions of the SW curves, 
using the geometry of NS 5-branes, D4-branes,
and O$6^{\pm}$ orientifold planes. 
How to compute subleading terms in elliptic models or 
their decompactification is one of the open problems of this subject.
Using these leading-order curves, 
we compute the one-instanton prepotential for each theory.
In section~5, we show that 
the SW curve obtained 
by Gukov and Kapustin \cite{GukovKapustin}
for SU$(N)$ with two antisymmetric hypermultiplets (with equal masses)
and four fundamental hypermultiplets,
and the curve
obtained by Uranga \cite{Uranga}
for SU$(N)$ with 
an antisymmetric and a symmetric hypermultiplet (with equal masses),
are equivalent, after a change of variables, 
to the curves for those theories derived in this paper,
giving dramatic confirmation of our methods. 
Section~6 is devoted to a 
consideration of SU$(N)$ gauge theory with a massive adjoint hypermultiplet.
We explicitly exhibit the relation between 
the curve derived in this paper,
the spectral curve derived from the Calogero-Moser model 
by D'Hoker and Phong \cite{DHokerPhong},
and the curve derived by Donagi and Witten \cite{DonagiWitten}
in the context of the integrable Hitchin system.
We close with conclusions and comments on open problems
in section~7.
 \vfil
\newpage


\noindent{\bf 2. ~Sp$(2N)\,+\,N_f$ fundamentals}
\renewcommand{\theequation}{2.\arabic{equation}}
\setcounter{equation}{0}

Although this paper primarily concerns elliptic models,
we include this section on 
Sp$(2N)$ gauge theory with $N_f$ fundamental hypermultiplets for completeness,
even though the SW curve is hyperelliptic. 
There are some unresolved 
issues when all the hypermultiplets in the fundamental representation
have non-zero masses \cite{DHokerKricheverPhong3}, 
and we take the opportunity to clarify these, particularly 
as the results are needed to complete our tables. 

The SW curve for Sp$(2N)$ gauge theory with $N_f<2N+2$ hypermultiplets
in the fundamental representation 
is \cite{LandsteinerLopezLowe,DHokerKricheverPhong3}
\be
y^2\,+\,2\, y \left[x^2\prod_{i=1}^N (x^2-e_i^2)\,+
\, i^{N_f}\, L^2 \prod_{j=1}^{N_f} M_j\right]\,+\,L^4 \prod_{j=1}^{N_f} 
(x^2-M_j^2)\,=\,0,
\label{dosuno}
\ee
where $L^2= \Lambda^{2N+2-N_f}$. 
Writing the curve (\ref{dosuno}) as 
\be
y^2\,+\,2\,A\,y\,+B=0,
\label{dosdos}
\ee
the SW differential is 
\be
\lambda={\left({A'\over A}-{B'\over 2B}\right)\,x \over 
\sqrt{1-B/A^2}}\,dx.
\label{dostres}
\ee
Because of the $x\rightarrow -x$ symmetry, and since the genus of the curve 
(\ref{dosuno}) exceeds the rank of the group (and hence the number of 
independent moduli), not all period integrals are relevant for the SW problem. 

The period integral is \cite{DHokerKricheverPhong3}
\be
2\pi i a_k\,=\, 2\,\int_{x^{-}_k}^{x^{+}_k}\, \lambda, 
\label{doscuatro}
\ee
as the $A_k$ cycles are taken to surround
the cut joining the two branch-points $x_k^{\pm}$.
The branch-cuts surrounding $x=e_k$ go from $x_k^{-}$ to 
$x_k^+$, and about $x=-e_k$ from $-x_k^{+}$ to 
$-x_k^{-}$. The $B_k$ cycle for the dual period 
is chosen to go from $-x_k^{-}$ to $x_k^{-}$ on the first sheet, 
and its counterpart on the second sheet. 
The dual period is given by  \cite{DHokerKricheverPhong3}
\be
2\pi i a_{D,k}\,=\, 2\,\int_{-x^{-}_k}^{x^{-}_k}\, \lambda. 
\label{doscinco}
\ee
The relevant branch-points are located at 
\be
x_k^{\pm}\,=\,e_k\,\pm\, \,L^2 \,[S_k(x_k^{\pm})]^{1/2}\,-\,L^2\, 
R_k(x_k^{\pm})
\,+\,{\cal O}(L^4),
\label{dosseis}
\ee
where 
\be
S_k(x)\,=\,{\prod_{j=1}^{N_f}\,(x^2-M_j{}^2)\over 
x^4\,(x+a_k)^2\,\prod_{i\not= k}^{N}\,(x^2-a_i{}^2)^2},
\label{dossiete}
\ee
and 
\be
R_k(x)\,=\,{i^{N_f}\,\prod_{j=1}^{N_f}\,M_j\over 
x^2\,(x+a_k)\,\prod_{i\not= k}^{N}\,(x^2-a_i{}^2)}.
\label{dosocho}
\ee

The periods and dual periods are computed by asymptotic expansion, 
as in refs.~\cite{DHokerKricheverPhong1,DHokerPhong}.
The period integral (\ref{doscuatro}) yields
\be
a_k\,=\,e_k\,-\,L^2\,R_k(e_k)\, +\,{L^4} 
\left( {\textstyle{1\over4}}\pr_k S_k(e_k) + \ha \pr_k\,[R_k(e_k)]^2\right)
\,+ \,{\cal O}(L^6).
\label{dosnueve}
\ee
Equation (\ref{dosnueve}) differs from the periods for 
SU$(N)$ \cite{DHokerKricheverPhong1} in that for Sp$(2N)$ 
$S_k(a_k)$ 
does not contribute to order $L^2$ (1-instanton). 
It will contribute to order $L^4$ (2-instanton), 
as is already clear from eq. (\ref{dosseis}) 
and will be explicitly shown below. 
To one-instanton accuracy, the dual periods are given by
\be
2\pi i a_{D,k}\,=\,2 \pi i (a_{D,k})_{\rm{classical}}\,+\,2 \pi i 
(a_{D,k})_{1-\rm{loop}}\, -
{8 L^2\over  a_k} \, \sum_{i=1}^{N}\,a_i\,R_i(a_i)\,+\,{\cal O}(L^4).
\label{dosdiez}
\ee
In order to integrate eq.  (\ref{dosdiez}) to obtain the one-instanton
prepotential, define the analytic function 
\be
F(x)\,=\,{i^{N_f}\,\prod_{j=1}^{N_f}\,M_j\over 
 x\,\prod_{i=1}^{N}\,(x^2-a_i{}^2)}.
\label{dosonce}
\ee
The sum of its residues vanishes, yielding
\be
2\,\sum_{i=1}^{N}\,a_i\,R_i(a_i)\,+\,{i^{N_f}\,\prod_{j=1}^{N_f}\,M_j\over 
\prod_{i=1}^{N}\,(-a_i{}^2)}=0.
\label{dosdoce}
\ee
With the definition
\be
S(x)\,=\,{\prod_{j=1}^{N_f}\,(x^2-M_j{}^2)\over 
x^4\,\prod_{j=1}^{N}\,(x^2-a_j{}^2)^2}\,=\,{\quarS_0(x)\over x^4},
\label{dostrece}
\ee 
the identity (\ref{dosdoce}) becomes
\be
2\,\sum_{i=1}^{N}\,a_i\,R_i(a_i)\,+\,[\quarS_0(0)]^{1/2}\,=\,0.
\label{doscatorce}
\ee
Since 
\be
{2\,[\quarS_0(0)]^{1/2}\over a_k}\,=\, - {\pr\over\pr a_k}[\quarS_0(0)]^{1/2},
\label{dosquince}
\ee
eq.~(\ref{dosdiez}) may be rewritten
\be
2 \pi i a_{D,k}\,=\,2 \pi i (a_{D,k})_{\rm{classical}}\,+\,
2 \pi i (a_{D,k})_{1-\rm{loop}}\, -
2 L^2{\pr\over \pr a_k} [\quarS_0(0)]^{1/2}\,+\,{\cal O}(L^4),
\label{dosdiezseis}
\ee
which can be integrated to give
\be
2\pi i {\cal F}_{\rm{1-inst}}\,=\,-2\,[\quarS_0(0)]^{1/2},
\label{dosdiezsiete}
\ee
with 
\be
[\quarS_0(0)]^{1/2}\,=\,i^{N_f}{\prod_{j=1}^{N_f}\,M_j \over 
\prod_{i=1}^{N}\,(-a_i{}^2)}.
\label{dosdiezocho}
\ee
This result is entered in Table 2. 
A similar derivation applies to Sp$(2N)$ + adjoint 
(the corresponding curve is given in sec.~4.6), and
to Sp$(2N)$ + 1  anti. + $N_f$ fund. (see sec.~4.7). 
The corresponding results are given in Table~2, with 
the relevant $S(x)$ given in Table~1. 

We can make several checks of our expressions 
(\ref{dosdiezsiete}) and (\ref{dosdiezocho}). 
First of all, for pure Sp$(2N)$ gauge theory, we have 
\be
2\pi i {\cal F}_{\rm{1-inst}}\,=\,-2(-1)^{N}\,
{1\over \prod_{i=1}^{N}\,a_i{}^2},
\label{dosdiezsietei}
\ee
which agrees with the results of
Masuda and Suzuki, and Ito and Sasakura \cite{japon} 
up to an overall constant, 
which can be reabsorbed with a redefinition of $\Lambda$.

Next, we can compare our one-instanton prediction 
for Sp(4) without matter hypermultiplets
with the one-instanton result 
for SO(5) without matter hypermultiplets
as given by \cite{DHokerKricheverPhong3,japon}. 
The order parameters $a_i$ of  Sp(4) are related to the order 
parameters $d_i$  of SO(5) by the change of variables
\be
a_1=(d_1+d_2)/2,\qquad a_2=(d_1-d_2)/2.
\label{thanksgiving} 
\ee
Inserting eq.~(\ref{thanksgiving}) into eq.~(\ref{dosdiezsietei}) 
we find again perfect agreement (up to an overall constant). 

We can also compare our result (\ref{dosdiezsiete}) 
for Sp(2) + $N_f$ fundamental hypermultiplets
with that for SU(2) + $N_f$ fundamentals 
as given for example in ref.~\cite{DHokerKricheverPhong1}:
\be
2\pi i {\cal F}_{\rm 1-inst}\,=\,
\cases{ 
{\Lambda_0{}^4\over 8  a^2}, ~~~~\qquad\qquad\qquad
\qquad\qquad\qquad\qquad\,\,\,\, {\rm for}\,\,\,\,\,
N_f=0,\cr 
{\Lambda_1{}^3\over 8 a^2}M_1, \qquad\qquad\qquad
\qquad\qquad\qquad\qquad\,\,\,\, {\rm for}\,\,\,\,\,
N_f=1,\cr 
{\Lambda_2{}^2\over 8 a^2}[a^2+M_1M_2], \qquad 
\qquad\qquad\qquad\qquad\,\,\,\,{\rm for}\,\,\,\,\,N_f=2,\cr
{\Lambda_3\over 8 a^2}[a^2(M_1+M_2+M_3)+M_1M_2M_3], 
\qquad {\rm for}\,\,\,\,\,
N_f=3.\cr
}
\label{su2}
\ee
Again, we find agreement up to a multiplicative constant, 
and a moduli-independent additive term.

At first glance, it appears that our result disagrees 
with the result for Sp($2N$) with $N_f$ fundamental hypermultiplets
given in ref.~\cite{DHokerKricheverPhong3}. 
In ref.~\cite{DHokerKricheverPhong3}, however,
at least two of the fundamental hypermultiplets had vanishing masses. 
In that case, 
eq. (\ref{dosdiezocho}) yields $[\quarS_0(0)]^{1/2}=0$,
and hence ${\cal F}_{\rm 1-inst}=0$ from eq.~(\ref{dosdiezsiete}). 
Thus, for the particular case of  Sp$(2N)$ with at least two massless 
fundamental hypermultiplets, 
the first non-trivial contribution to the instanton prepotential 
is ${\cal F}_{\rm 2-inst}$.

To make contact with the results in ref.~\cite{DHokerKricheverPhong3},  
we calculate the two-instanton contribution to the prepotential
following the method of ref. \cite{Chan}. 
The result is 
\be
2\pi i {\cal F}_{\rm{2-inst}}\,=\,\sum_{k=1}^N S_k(a_k)\,+\,{1\over 4}
\left({\pr^2 \,{\quarS_0}\over\pr x^2} \right)_{x=0}.
\label{twoins}
\ee
Note that $S_k(a_k)$ contributes to two instantons 
(as it depends on $L^4$), as we had anticipated.  
Further, from eq.~(\ref{dostrece}),
one may verify that
${\pr^2 \,{\quarS_0}\over \pr x^2} (0) = 0$
when two or more of the hypermultiplets are massless,
so the only contribution to the two-instanton prepotential  
will be the first term in eq.~(\ref{twoins}),
in complete agreement with \cite{DHokerKricheverPhong3}. 

For generic values of the masses of the matter hypermultiplets, 
we can check our two-instanton result for  Sp(2) 
against the two-instanton result for SU(2), 
given by the expressions~\cite{DHokerKricheverPhong1}:
\be
2\pi i {\cal F}_{\rm 2-inst}\,=\,
\cases{ 
{5\Lambda_0{}^8\over 2^{10} a^6}, \qquad\qquad\qquad\qquad\qquad\qquad
\qquad\qquad\qquad\qquad\qquad\,\,\,\,{\rm for}\,\,\,\,\,\,
N_f=0,\cr 
{\Lambda_1{}^6\over 2^{10} a^6}\left[5M_1^2-3a^2\right], \qquad\qquad\qquad
\qquad\qquad\qquad\qquad\qquad\,\,\,\, \,{\rm for}\,\,\,\,\,\,
N_f=1,\cr 
{\Lambda_2{}^4\over 2^{10} a^6}\left[a^4-3a^2(M_1^2+M_2^2)+5M_1^2M_2^2\right], 
\qquad\qquad\qquad\qquad\,{\rm for}\,\,\,\,\, N_f=2,\cr
{\Lambda_3^2\over 2^{10} a^6}\left[a^6+a^4(M_1^2+M_2^2+M_3^2)\right.\cr
\left.\qquad\!\!\!-3a^2(M_1^2M_2^2+M_2^2M_3^2+M_1^2M_3^2)+5M_1^2M_2^2M_3^2\right], 
\qquad {\rm for}\,\,\,\,\,
N_f=3.\cr
}\nn\\
\label{su2dos}
\ee
Our results (\ref{twoins}) agree with eq. (\ref{su2dos}) 
up to an overall constant. 
(For $N_f=3$ there is a moduli-independent additive 
constant as well.)

Finally, we can compare our two-instanton prediction (\ref{twoins})
for Sp(4) without matter hypermultiplets
with the two-instanton result for SO(5) 
obtained using the method of ref. \cite{Chan}. 
Using the change of variables (\ref{thanksgiving})
we again find agreement.\footnote{Apparently,
there is a misprint 
in eq. (3.9) for the two-instanton prepotential for SO($N$) 
in ref.~\cite{DHokerKricheverPhong3}. The correct expression appears to be 
\be
{\cal F}_{\rm 2-inst} \propto \sum_{k\not= l}^{N} 
{S_k(a_k)S_l(a_l)\over (a_k-a_l)^2}
+\sum_{k,l}^{N} {S_k(a_k)S_l(a_l)\over (a_k+a_l)^2}+{1\over4}\sum_{k=1}^{N}S_k(a_k)
{\pr^2 S_k(x)\over \pr x^2}\Big\vert_{x=a_k}.
\label{mierda}
\ee}

\vskip.3in

\noindent{\bf 3. ~Universality}
\renewcommand{\theequation}{3.\arabic{equation}}
\setcounter{equation}{0}

By examining ${\cal F}_{\rm 1-inst}$ 
obtained for all generic cases of classical groups in the Coulomb phase, 
one finds that the results may be summarized succinctly
in terms of a master function $S(x)$ for each case, 
as we have emphasized previously \cite{twoanti,santiago}.
These functions are collected for each theory in  Table~1.
(The new results in this table are from sections~2~and~4 of this paper.)

\begin{center} 
\begin{tabular}{||c|c||} 
\hline
\hline Hypermultiplet
Representations & $S(x)$\\ 
\hline\hline 
SU$(N)\,+\,N_f \, {\rm fund.}\, (M_j)$ &{} \\ 
$(N_f \leq 2N)$ 
& ${\xf \prod_{j=1}^{N_f}(x+M_j)\over \prod_{i=1}^N (x-a_i)^2}$    \\ 
(ref. \cite{DHokerKricheverPhong1}) &{} \\ 
\hline
${\rm SU}(N)\,+\,1\,\,{\rm sym.}\,(m)  \,+\, N_f\,{\rm fund.} \, (M_j) $ &{}\\
$(N_f \,\leq N-2) $ 
& $ {\xf (-1)^N(x+\nha m)^2 \prod_{i=1}^N (x+a_i+\xxt m)
\prod_{j=1}^{N_f}(x+M_j) \over \prod_{i=1}^N (x-a_i)^2}$         \\
(ref. \cite{onesym,nonhyper}) &{} \\ 
\hline 
${\rm SU}(N)\,+\,1\,\,{\rm anti.} (m) \,+\, N_f\,{\rm fund.}\,(M_j)  $ &{} \\ 
$(N_f\,\,\leq N+2)$ &
${\xf (-1)^N \prod_{i=1}^N (x+a_i+\xxt m) \prod_{j=1}^{N_f}(x+M_j) \over
(x+\nha m)^2 \prod_{i=1}^N (x-a_i)^2}$ \\ 
(ref. \cite{oneanti,nonhyper}) & \\ 
\hline 
${\rm SU}(N)\,
+\,2 \,\,{\rm anti.} \,(m_1,m_2) \,+\, N_f\,{\rm fund.}\,(M_j)  $ &{} \\ 
$(N_f \,\,\leq 4)$ & ${\xf \prod_{i=1}^{N}(x+a_i+\xxt m_1) \prod_{i=1}^N(x
+a_i+\xxt m_2) \prod_{j=1}^{N_f}(x +M_j) \over (x +\nha m_1)^2 (x +\nha m_2)^2
\prod_{i=1}^N (x -a_i)^2}$ \\ (ref. \cite{twoanti}) 
&{} \\ \hline
 &{} \\
SU$(N)$ + 1  anti. ($\xxt m_1$) + 1 sym. ($\xxt m_2$)  & 
${\xf  (x+\nha m_2)^2\,
\prod_{i=1}^N (x+a_i+\xxt m_1)  \prod_{i=1}^N (x+a_i+\xxt m_2) \over
(x+\nha m_1)^2 \prod_{i=1}^N (x-a_i)^2}$ \\ 
 & \\ 
\hline  &{} \\
${\rm SU}(N)$ + adjoint & 
${\xf \prod_{i=1}^N [(x-a_i)^2-m^2] \over\prod_{i=1}^N (x-a_i)^2}$ \\ 
(ref.  \cite{DHokerPhong}) & \\ 
\hline \end{tabular} \end{center} \label{tableone}
\centerline{\footnotesize{\bf Table 1}}

\vspace{1.5cm}

\begin{center} 
\begin{tabular}{||c|c||} 
\hline
\hline Hypermultiplet
Representations & $S(x)$\\ 
\hline
\hline 
 SO$(2N) \, + \, N_f\, {\rm fund.}  $ &  {} \\
 $(N_f \leq 2N-2)$ & 
${\xf  x^4 \prod_{j=1}^{N_f}(x^2-M_j^2)\over \prod_{i=1}^N (x^2-a_i^2)^2}$ \\ 
(ref. \cite{DHokerKricheverPhong3})  &{} \\ 
\hline 
&{} \\ 
SO$(2N)+ {\rm adjoint}$ & 
$ {\xf  x^4 \prod_{i=1}^N [(x-m)^2 -a_i^2] \prod_{i=1}^N [(x+m)^2 -a_i^2]
\over { (x+\ha m)^2 (x-\ha m)^2 \prod_{i=1}^N (x^2-a_i^2)^2}}$       \\
{} &{} \\ 
\hline 
 SO$(2N+1)\,+\, N_f\, {\rm fund.}$ &  {} \\
 $(N_f \leq 2N-1)$ & ${\xf  x^2 \prod_{j=1}^{N_f}(x^2-M_j^2)\over
\prod_{i=1}^N (x^2-a_i^2)^2}$    \\ 
(ref. \cite{DHokerKricheverPhong3})  &{} \\ 
\hline
 &{} \\ 
SO$(2N+1)+ {\rm adjoint}$ & 
$ {\xf  x^2 (x+m)(x-m) \prod_{i=1}^N [(x-m)^2 -a_i^2] 
\prod_{i=1}^N [(x+m)^2 -a_i^2]
\over { (x+\ha m)^2 (x-\ha m)^2 \prod_{i=1}^N (x^2-a_i^2)^2}}$       \\
{} &{} \\ 
\hline 
 Sp$(2N) \,+\, N_f\, {\rm fund.}$ &  {} \\
 $(N_f \leq 2N+2)$ & ${\xf  \prod_{j=1}^{N_f}(x^2-M_j^2)\over
{ x^4  \prod_{i=1}^N (x^2-a_i^2)^2}}$    \\ 
{} &{} \\ 
\hline
 &{} \\ 
Sp$(2N)+ {\rm adjoint}$ & 
$ {\xf  (x+\ha m)^2 (x-\ha m)^2 \prod_{i=1}^N [(x-m)^2 -a_i^2]
\prod_{i=1}^N [(x+m)^2 -a_i^2]
\over { x^4\prod_{i=1}^N (x^2-a_i^2)^2}}$       \\
{} &{} \\ 
\hline 
Sp$(2N)\, + 1\,{\rm anti.}\,+\,N_f\, {\rm fund.}$ 
&{} \\ 
$(N_f \leq 4) $
& $ {\xf  \prod_{i=1}^N [(x-m)^2 -a_i^2] \prod_{i=1}^N [(x+m)^2 -a_i^2]
\prod_{j=1}^{N_f}(x^2-M_j^2)
\over { x^4 (x+\ha m)^2 (x-\ha m)^2  \prod_{i=1}^N (x^2-a_i^2)^2}}$       \\
&{} \\ 
\hline 
\hline 
\end{tabular} 
\end{center} 
\label{tablebis}
\centerline{\footnotesize{\bf Table 1}: (Continuation).}
\vspace{1.5cm}

Given $S(x)$, one defines residue functions 
$S_k(x)$ and $S_m(x)$ at the quadratic poles of $S(x)$ by 
\be
S(x)\,=\,{S_k(x)\over (x-a_k)^2}\, =\,{S_m(x)\over (x+\nha m)^2}.
\label{dosone} 
\ee 
If $S(x)$ has a quartic pole at $x=0$, one defines
\be
S(x)\,=\, {\quarS_0(x) \over x^4}.
\label{dosquartic} 
\ee 
In many cases \cite{onesym,nonhyper,DHokerKricheverPhong1,DHokerKricheverPhong3,
DHokerPhong},
the one-instanton prepotential is given by
\be 
\xe \pi i {\cal F}_{\rm
1-inst}\,=\,\sum_{k=1}^N S_k(a_k), 
\label{dostwo} 
\ee 
while for models containing one antisymmetric representation of 
SU$(N)$ \cite{oneanti,nonhyper}
or the adjoint representation of SO($N$), the one-instanton prepotential is
\be 
\xe \pi i {\cal F}_{\rm
1-inst}=\sum_{k=1}^{N} S_k(a_k)-2S_m(-\nha m), 
\label{dosthree} 
\ee 
and for models containing two antisymmetric representations of 
SU$(N)$ \cite{twoanti}, it is
\be 
\xe \pi i {\cal F}_{\rm 1-inst}=
\sum_{k=1}^N S_k(a_k) -2 S_{m_1}(-\nha m_1)-2S_{m_2}(-\nha m_2).
\label{dosfour} 
\ee 
Finally for Sp$(2N)$ with various matter content, 
the one-instanton prepotential is 
\be 
\xe \pi i {\cal F}_{\rm 1-inst}=- 2 [\quarS_0(0)]^{1/2}.
\label{zdosfour}
\ee 
These results are summarized in Table~2.

\begin{center}
\begin{tabular}{||c|c||}
\hline\hline
Group & Matter content\\
\hline\hline
\multicolumn{2}{||c||}
{ $\xe \pi i{\cal F}_{\rm 1-inst}\,= \, \sum_{k=1}^N S_k(a_k)$}\\
\hline
{}     & $N_f$ ~fund. ($N_f\leq 2N$) \\ \cline{2-2}
${\rm SU}(N)$       & 1 sym. + $N_f$ ~fund.  ($N_f \leq N-2$) \\ \cline{2-2}
                   & adjoint                    \\
\hline
SO$(2N)$  & $N_f$ ~fund. ($N_f \leq 2N-2$)\\ 
\hline
SO$(2N+1)$  & $N_f$ ~fund. ($N_f\leq 2N-1$)\\ 
\hline
\hline
\multicolumn{2}{||c||}{ 
$\xe \pi i{\cal F}_{\rm 1-inst}\,=\, 
\sum_{k=1}^N S_k(a_k)-2S_{m_1}(-\nha m_1)$}\\
\hline
${\rm SU}(N)$ & 1 anti. ($\xxt m_1$) + $N_f$ ~fund.  ($N_f\leq N+2$)\\ 
\cline{2-2}
              & 1 anti. ($\xxt m_1$) + 1 sym. ($\xxt m_2$)\\
\hline   
SO$(2N)$  & adjoint ($\xxt m_1$) \\ 
\hline
SO$(2N+1)$  & adjoint ($\xxt m_1$) \\ 
\hline  
\hline     
\multicolumn{2}{||c||}{ 
$\xe \pi i{\cal F}_{\rm 1-inst}\,=\, 
\sum_{k=1}^N S_k(a_k)-2S_{m_1}(-\nha m_1)-2S_{m_2}(-\nha m_2)$}\\
\hline
${\rm SU}(N)$  & 2 anti. ($\xxt m_1,\xxt m_2$)  + $N_f$ ~fund. ($N_f \leq 4$)\\  
\hline     
\hline
\multicolumn{2}{||c||}{ $\xe \pi i{\cal F}_{\rm 1-inst}\,=\, 
-2 [\quarS_{0}(0)]^{1/2}$}\\
\hline
                  &  $N_f$ ~fund. ($N_f \leq 2N+2$) \\ \cline{2-2}  
Sp$(2N)$          & adjoint \\ \cline{2-2}  
                  & 1 anti. +~$N_f $ ~fund. ($N_f \leq 4$) \\
\hline		 			 
\hline
\end{tabular}
\end{center}
\label{tabletwo}
\centerline{{\footnotesize{\bf Table 2}: ${\cal F}_{\rm 1-inst}$ 
for different groups and  matter content.}}

\begin{center} 
\begin{tabular}{||c|c|c||} 
\hline
\hline 
Group & Representation & 
Factor of  $S(x)$\\ 
\hline
\hline 
{} & gauge multiplet & ${\xf 1 \over {\prod_{i=1}^N (x-a_i)^2}}$ \\
\cline{2-3}
{} & $N_f$ fundamental ($M_j$)  & $ \prod_{j=1}^{N_f} (x+M_j)$  \\
\cline{2-3}
${\rm SU}(N)$ & symmetric ($\xxt m$) & 
$(-1)^N (x+\nha m)^2 \prod_{i=1}^N (x+a_i+\xxt m)$ \\
\cline{2-3}
{} & antisymmetric  ($\xxt m$)  &  
$ (-1)^N (x+\nha m)^{-2} \prod_{i=1}^N (x+a_i+\xxt m)$ \\
\cline{2-3}
{} & adjoint ($m$) & $\prod_{i=1}^N [(x-a_i)^2-m^2]$\\
\hline
\hline
{} & gauge multiplet & ${\xf x^4 \over {\prod_{i=1}^N (x^2-a_i^2)^2}}$ \\
\cline{2-3}
SO$(2N)$ & $N_f$ fundamental ($M_j$)  & $ \prod_{j=1}^{N_f} (x^2-M_j^2)$  \\
\cline{2-3}
{} & adjoint ($m$) &  ${\prod_{i=1}^N [(x+m)^2-a_i^2]
\prod_{i=1}^N [(x-m)^2-a_i^2] \over {(x+\ha m)^2(x-\ha m)^2}}$ \\
\hline
\hline
{} & gauge multiplet & ${\xf x^2 \over {\prod_{i=1}^N (x^2-a_i^2)^2}}$ \\
\cline{2-3}
SO$(2N+1)$ & $N_f$ fundamental ($M_j$)  & $ \prod_{j=1}^{N_f} (x^2-M_j^2)$  \\
\cline{2-3}
{} & adjoint ($m$) &  ${(x+m)(x-m) \prod_{i=1}^N [(x+m)^2-a_i^2]
\prod_{i=1}^N [(x-m)^2-a_i^2] \over {(x+\ha m)^2(x-\ha m)^2}}$ \\
\hline
\hline
{} & gauge multiplet & ${1 \xf  \over {x^4 \prod_{i=1}^N (x^2-a_i^2)^2}}$ \\
\cline{2-3}
Sp$(2N)$ & $N_f$ fundamental ($M_j$)  & 
$ \prod_{j=1}^{N_f} (x^2-M_j^2)$  \\
\cline{2-3}
{} & adjoint ($m$) &  $ (x+\ha m)^2 (x-\ha m)^2 \prod_{i=1}^N [(x+m)^2-a_i^2]
\prod_{i=1}^N [(x-m)^2-a_i^2]$ \\
\cline{2-3}
{} & antisymmetric ($m$) &  $ {\prod_{i=1}^N [(x+m)^2-a_i^2] 
\prod_{i=1}^N [(x-m)^2-a_i^2] \over {(x+\ha m)^2 (x-\ha m)^2}}$\\
\hline
\hline
\end{tabular} 
\end{center} 
\label{tablethree}
\centerline{\footnotesize{\bf Table 3}: Factors of $S(x)$.}
\vspace{0.3cm}
\begin{center} 
\begin{tabular}{||c|c||} 
\hline
\hline 
   ${\rm SU}(N)$ +   1 sym. + 1 anti. 
&  ${\rm SU}(N)$ +  2 anti. + 4 fund. \\
$(m,m)$ & $(m,m) + (\nha m,\nha m,\nha m,\nha m)$ \\
\hline
\multicolumn{2}{||c||}{Valid for all moduli.} \\
\multicolumn{2}{||c||}{Also verified using elliptic curves
(refs.~\cite{Uranga} and \cite{GukovKapustin}). See section 5.}\\
\hline
\hline 
SO$(2N)$ + adjoint  &  Sp$(2N)$ + 1 anti. + 4 fund.\\
$(m)$  & $(m) + (0,0,0,0)$\\
\hline
\multicolumn{2}{||c||}{Valid for all moduli.}\\
\hline
\hline 
Sp$(2N)$  + adjoint &  Sp$(2N)$ + 1 anti + 4 fund. \\
$(m)$  & $(m) + (\nha m,\nha m,\nha m,\nha m)$ \\
\hline
\multicolumn{2}{||c||}{Valid for all moduli}\\
\hline
\hline 
SU$(2N)$ +  2 anti. + 4 fund.  &  SO$(2N)$ + adjoint \\
$(m,-m)\,+\, (0,0,0,0)$ &  $\,\,\,\,\,\,\,\,\,(m)$ \\
Moduli:          & Moduli: \\
$a_1,a_2,\cdots,a_N,-a_1,-a_2, \cdots,-a_N $  & $a_1,a_2, \cdots,a_N$\\
\hline
\hline
SU$(2N)$ + 2 anti. + 4 fund.  &  Sp$(2N)$ + 1 anti. + 4 fund. \\
$(m,-m)\,+\, (M_1,M_2,-M_1,-M_2)$  & $(m)\,+\, (M_1,M_2,0,0)$\\
\hline
\multicolumn{2}{||c||}{Same relations for moduli as in the previous case.}\\
\hline
\hline
\end{tabular} 
\end{center} 
\label{tablefour}
\centerline{\footnotesize{\bf Table 4}: Table of equivalences (from $S(x)$).}
\vspace{1.5cm}

An examination of Table 1 leads to empirical rules for 
constructing $S(x)$, where $S(x)$ is given as the product of factors, 
each corresponding to a different ${\cal N}=2$ multiplet in 
a given representation of a classical group. 
These rules for the factors that make up $S(x)$ are
given in Table 3, 
which contains some new results, 
obtained in secs.~2 and 4 of this paper.

By examining Table 3, one observes that certain pairs of
(mass-deformed) elliptic models have identical $S(x)$,
and therefore identical ${\cal F}_{\rm 1-inst}$,
for suitable choices of mass parameters and moduli. 
These equivalences are presented in Table 4. 
In each of the cases, one may verify that
${\cal F}_{\rm 1-loop}$ is also identical for both sides.\footnote{This
has been previously observed for the third entry of the table
in ref.~\cite{DLS}.}
Finally, 
one can verify that the curves are identical 
on both sides of the first line of the table, 
and that the leading-order terms (see section 4)
of the curves are identical
on both sides of the remaining lines of the table.
Since it is very plausible
that $S(x)$ determines the complete instanton expansion for a theory,
we claim that Table 4 likely represents pairs of theories with 
identical prepotentials. 

\vskip.3in
\noindent{\bf 4. ~Curves for decompactified elliptic models.}
\renewcommand{\theequation}{4.\arabic{equation}}
\setcounter{equation}{0} 

In secs.~4 and 5 of ref. \cite{twoanti}, 
we reverse-engineered a curve for 
${\cal N} = 2$  SU$(N)$ gauge theory
with two hypermultiplets in the antisymmetric representation 
and $N_f < 4$ hypermultiplets in the fundamental representation
using the regularities of the function $S(x)$ observed in
sec.~3 of this paper.
This curve can be regarded as the decompactification 
of an elliptic model with zero beta function, 
having two hypermultiplets in the antisymmetric representation 
and four fundamental hypermultiplets, 
with the mass of one or more of the fundamental representations
sent to infinity. 
The resulting SW curve remains of infinite order in this limit.

A number of ${\cal N} = 2$ theories with simple classical gauge groups 
can be understood as decompactifications of elliptic models.
(By decompactification, we mean that the curve is formulated 
on the covering space of the circular base space of the elliptic model.)
The M-theory description for these theories 
has been considered by Uranga \cite{Uranga}, 
but only for those with zero beta function.
The ``basic cell" for 
these models (except for SU$(N)$ with an adjoint hypermultiplet)
contains two O$6$ planes
(with the same or opposite charges depending on the group and matter content), 
together with one or two NS 5-branes, 
and a number of D4-branes and (possibly) D6-branes. 
(We use the language of Type~IIA theory, 
which is then considered to be lifted to M-theory.) 
In most cases, no explicit curve, 
suitable for studying the instanton expansion,
has been previously presented 
(except for SU$(N)$ with a massive adjoint hypermultiplet, 
for which the prepotential was calculated in ref.~\cite{DHokerPhong}).
In this section, we present curves and 
the resulting prepotentials
for those theories obtained as 
decompactifications of the elliptic models discussed in \cite{Uranga}. 

The curve for a decompactified elliptic model has the form
\be
\sum_{n=-\infty}^\infty L^{\arh n^2}\, J_n(x)\, P_n(x)\,t^n\, =0, 
\label{infcurve}
\ee
where $t= \exp[-(x_6 + i x_{10})/R]$, $x = x_4 + i x_5$, 
and $\arh = 1$ ($\arh = 2$) if there are two (one) NS 5-branes
per unit cell. 
The coefficient functions $P_n(x)$ and $J_n(x)$ themselves have 
(in principle) expansions in $L$,
\be
P_n(x)  & = &  P_n(x) \Big|_{\rm leading} +  {\cal O} (L^2)\,, \nn\\
J_n(x)  & = &  J_n(x) \Big|_{\rm leading} +  {\cal O} (L^2)\,. 
\label{leading}
\ee 
The leading-order terms in $P_n(x)$ are determined by the positions 
of the D4-branes ({\ie}, moduli of the gauge theory)
and the positions of the orientifold planes ({\ie}, masses of hypermultiplets 
in two-index representations of the gauge group),
while the leading-order terms of $J_n(x)$ are associated with 
the positions of the D6-branes ({\ie}, masses of fundamental representations),
if present.
We have not been able to uniquely determine the subleading terms
in eq.~(\ref{leading}).

The M-theory pictures corresponding to 
elliptic models with zero beta function
are periodic in both the $x_6$ and $x_{10}$ directions.
If we let $z$ parametrize the torus 
with the identifications $z \equiv z + 2 \omega_1 \equiv z + 2 \omega_2$,
then $t=e^{\beta z}$ with  $\beta = -i \pi/\omega_1$.
The shift $z \to z + 2\omega_1$ (or $t \to e^{-2 \pi i} t$)
corresponds to a translation by $2 \pi R$ in the $x_{10}$ direction.
The shift $z \to z + 2\omega_2$ (or $t \to q^{-1} t$),
where $q = e^{2\pi i\tau}$ and $\tau = \omega_2/\omega_1$,
corresponds to a translation in the $x_6$ direction 
(accompanied by a translation in $x_{10}$ if Re $\tau \neq 0$).
The curves (\ref{infcurve}) derived from these pictures 
would likewise be expected to be doubly periodic (up to a shift in $x$).
Periodicity in the $x_{10}$ direction is automatic,
but periodicity in the $x_6$ direction
requires that $L$ be replaced by $q^{1/4}$, so (\ref{infcurve}) becomes
\be
\sum_{n=-\infty}^\infty q^{\arh n^2/4}\,e^{\beta n z}\, J_n(x)\, P_n(x)\, =0, 
\label{ellipcurve}
\ee
for theories with zero beta function.
It further requires that $P_n(x)$ and $J_n(x)$ 
possess certain symmetry properties.
If there are two NS 5-branes per unit cell 
(as is the case for the SU$(N)$ theories with 
symmetric or antisymmetric hypermultiplets that we consider), 
then
\be
P_{n + 2\ell }(x)&=&P_n (x-\ell\Delta),\nn\\
J_{n + 2\ell }(x)&=&J_n (x-\ell\Delta),
\label{twobranes}
\ee
implies that the curves (\ref{infcurve}) and (\ref{ellipcurve}), 
with $\arh=1$, are invariant under 
$ t \to t L^{-4}$ (or $z \to z + 2\omega_2$)
and $ x \to x + \Delta $
(where $\Delta$ is the ``global mass,'' 
the relative mass of the two hypermultiplets in two-index 
representations of SU($N$)).
If there is only one NS 5-brane per unit cell 
(as in all the SO$(N)$ and Sp$(2N)$ theories that we consider,
or the SU$(N)$ theory with a massive adjoint hypermultiplet), 
then
\be
P_{n + \ell}(x)&=&P_n (x-\ell m),\nn\\
J_{n + \ell}(x)&=&J_n (x-\ell m),
\label{onebrane}
\ee
guarantees that the curves (\ref{infcurve}) and (\ref{ellipcurve}), 
with $\arh = 2$, 
are invariant under 
$ t \to t L^{-4}$ (or $z \to z + 2\omega_2$)
and $ x \to x + m $ 
(where $m$ is the global mass, 
the mass of the adjoint or antisymmetric hypermultiplet).

The factors in the leading terms of $P_n(x)$ and $J_n(x)$ correspond
not only to the D4-branes which depend on the moduli,
but also to the semi-infinite ``non-dynamical'' D4-branes associated
with the O6 planes and D6-branes \cite{Uranga,HananyWitten}.
The placement of these non-dynamical D4-branes is not unique,
because they can extend either to the left or the right of
the O6 plane or D6-brane.
Different choices correspond precisely
to different parametrizations of the curve $t \to t/G(x) $,
where $G(x)$ is a rational function of $x$ and positions of
O6 planes and D6-branes.
The symmetries (\ref{twobranes}) or (\ref{onebrane}) will
not be present unless the non-dynamical D4-branes are distributed
symmetrically to the left and the right.
In this paper, therefore, we will always 
choose a parametrization of the curves 
that respects these symmetries,
so that the invariance of the curve under translations in $x_6$
(together with a shift in $x$) will be manifest.

To determine the prepotential to 1-instanton accuracy, 
it is sufficient \cite{product}--\cite{twoanti}
to consider only the quartic truncation of the curve (\ref{infcurve})
\be
L^{4\arh}\,J_2(x)\,P_2(x)\,t^2
&+&L^{\arh}  \,J_1(x)\,P_1(x)\,t  \,
+ \,     J_0(x)\,P_0(x)\,\nn\\
&+& L^{\arh}  \,J_{-1}(x)\,P_{-1}(x)\,{1\over t}\,
+\, L^{4\arh}\,J_{-2}(x)\,P_{-2}(x)\,{1\over t^2}=0.
\label{trunccurve}
\ee
The prepotential for the decompactified elliptic models 
may then be obtained by calculating the period integrals $a_k$ and $a_{D,k}$
from the curve (\ref{infcurve}), 
and then integrating
$ a_{D,k}={\pr {\cal F}/ {\pr a_k}}$.
Applying residue methods \cite{DHokerKricheverPhong1}
and hyperelliptic perturbation theory \cite{oneanti}--\cite{twoanti}
to the quartic truncation (\ref{trunccurve}),
one obtains\footnote{For the Sp$(N)$ theories, 
however, the one instanton prepotential is proportional
to $L^2$, not $L^{2 \arh}$.}
\be
{\cal F}(A)\,=\, {\cal F}_{\rm classical}(A)\, 
+\,{\cal F}_{\rm 1-loop}(A)\,+\,L^{2\arh} {\cal F}_{\rm 1-inst}(A), 
\ee
where ${\cal F}_{\rm 1-inst}(A)$
is given by 
eqs.~(\ref{dostwo}), (\ref{dosthree}), (\ref{dosfour}), or (\ref{zdosfour}).
In each case, the function $S(x)$ is given by
\be
S(x)
= \left. \xf {J_1(x)\, J_{-1}(x)\,P_1(x)\, P_{-1}(x)\over J_0^2(x)\,P_0^2(x)}
\right|_{\rm leading} ,
\label{masterfcn}
\ee
where only the leading-order terms 
of $P_n(x)$ and $J_n(x)$ are used in defining $S(x)$.
If there are no D6-branes in the model,
the factors of $J_n(x)$ are absent in  (\ref{masterfcn}).
The $O(L^{4\arh})$ terms in  (\ref{trunccurve}) are essential in obtaining
the one-instanton prepotential, 
which involves a delicate cancellation~\cite{oneanti}
between these terms and the subleading terms in $P_n(x)$.
This cancellation places constraints on the form of these subleading
terms, but does not necessarily uniquely determine them.

\vfil
\eject
In the rest of this section, we present results for
each of several models, giving in each case:

\noindent
a) a figure containing the M-theory picture of the model;

\noindent
b) the leading-order terms of the coefficient functions $P_n(x)$ and $J_n(x)$,
{\ie}, those terms with the lowest power in 
$L$ or $q$ for a given $n$; 

\noindent
c) the infinite curve for each model (with leading-order coefficients only);

\noindent
d) the function $S(x)$ calculated from each curve;

\noindent
e) the one-instanton contribution to the prepotential;  

\noindent
f) various checks on the proposed curve and prepotential.

\vfil
\eject

\noindent{\bf 4.1.}
\underline{ SU$(N)\,+\,2\, {\rm antisymmetric}\,+\,N_f\,\,{\rm fundamentals}$ }
\renewcommand{\theequation}{4.1.\arabic{equation}}
\setcounter{equation}{0}

Consider SU$(N)$ gauge theory 
with two matter hypermultiplets (masses $m_1$ and $m_2$)
in the antisymmetric representation,
and $N_f$ matter hypermultiplets (masses $M_j$)
in the fundamental (defining) representation.
The M-theory picture for this case, represented in Fig. 1,
contains an infinite chain of NS 5-branes,
with an O$6^{-}$ plane coincident with each one.
Between each pair of consecutive 5-branes there are $N$ D4-branes  
and $N_f$ D6-branes.

\begin{center}
\begin{picture}(810,295)(10,10)


\put(10,192){\line(1,0){5}}
\put(20,192){\line(1,0){5}}
\put(30,192){\line(1,0){5}}
\put(40,192){\line(1,0){5}}
\put(50,192){\line(1,0){5}}
\put(60,192){\line(1,0){5}}
\put(70,192){\line(1,0){5}}
\put(80,192){\line(1,0){5}}
\put(90,192){\line(1,0){5}}
\put(100,192){\line(1,0){5}}
\put(5,180){$(\!x\!-\!a_i\!+\!\xtw m_2\!-\!\xtw m_1\!)$}

\put(105,240){\line(1,0){2}}
\put(110,240){\line(1,0){5}}
\put(120,240){\line(1,0){5}}
\put(130,240){\line(1,0){5}}
\put(140,240){\line(1,0){5}}
\put(150,240){\line(1,0){5}}
\put(160,240){\line(1,0){5}}
\put(170,240){\line(1,0){5}}
\put(180,240){\line(1,0){5}}
\put(190,240){\line(1,0){5}}
\put(200,240){\line(1,0){5}}
\put(120,245){$(\!x\!+\!a_i\!+\!\xtw m_2\!)$}

\put(205,117){\line(1,0){2}}
\put(210,117){\line(1,0){5}}
\put(220,117){\line(1,0){5}}
\put(230,117){\line(1,0){5}}
\put(240,117){\line(1,0){5}}
\put(250,117){\line(1,0){5}}
\put(260,117){\line(1,0){5}}
\put(270,117){\line(1,0){5}}
\put(280,117){\line(1,0){5}}
\put(290,117){\line(1,0){5}}
\put(300,117){\line(1,0){5}}
\put(235,105){$(\!x-\!a_i\!)$}

\put(305,155){\line(1,0){2}}
\put(310,155){\line(1,0){5}}
\put(320,155){\line(1,0){5}}
\put(330,155){\line(1,0){5}}
\put(340,155){\line(1,0){5}}
\put(350,155){\line(1,0){5}}
\put(360,155){\line(1,0){5}}
\put(370,155){\line(1,0){5}}
\put(380,155){\line(1,0){5}}
\put(390,155){\line(1,0){5}}
\put(400,155){\line(1,0){5}}
\put(320,160){$(\!x\!+\!a_i\!+\!\xtw m_1\!)$}

\put(405,10){\line(1,0){2}}
\put(410,10){\line(1,0){5}}
\put(420,10){\line(1,0){5}}
\put(430,10){\line(1,0){5}}
\put(440,10){\line(1,0){5}}
\put(450,10){\line(1,0){5}}
\put(460,10){\line(1,0){5}}
\put(470,10){\line(1,0){5}}
\put(480,10){\line(1,0){5}}
\put(490,10){\line(1,0){5}}
\put(500,10){\line(1,0){5}}
\put(407,15){$(\!x\!-\!a_i\!+\!\xtw m_1\!-\!\xtw m_2\!)$}

\put(50,-10){$P_{2}$}
\put(150,-10){$P_{1}$}
\put(250,-10){$P_0$}
\put(350,-10){$P_{-1}$}
\put(450,-10){$P_{-2}$}


\put(105,5){\line(0,1){255}}
\put(205,5){\line(0,1){255}}
\put(305,5){\line(0,1){255}}
\put(405,5){\line(0,1){255}}
\put(103,271){$\NSone$}
\put(203,271){$\NStwo$}
\put(303,271){$\NSthree$}
\put(403,271){$\NSfour$}


\put(101,215){$\otimes$} 
\put(201,175){$\otimes$}
\put(301,135){$\otimes$}
\put(401,85){$\otimes$}
\put(31,220){$(\!x\!+\!\xxt m_2\!-\!\nha m_1\!)$}
\put(160,170){$(\!x\!+\!\nha m_2\!)$}
\put(259,135){$(\!x\!+\!\nha m_1\!)$}
\put(406,75){$(\!x\!+\!\xxt m_1\!-\!\nha m_2\!)$}

\put(110,220){O$6^{-}$}
\put(210,180){O$6^{-}$}
\put(310,140){O$6^{-}$}
\put(410,90){O$6^{-}$}

\put(151,185){\framebox(5,5){$\cdot$}}
\put(251,165){\framebox(5,5){$\cdot$}}
\put(351,105){\framebox(5,5){$\cdot$}}
\put(120,194){$(\!x\!+\!\xxt m_2\!-\!M_j\!)$}
\put(235,172){$(\!x\!+\!M_j\!)$}
\put(320,96){$(\!x+\!\xxt m_1\!-\!M_j\!)$}

\put(30,30){\vector(1,0){20}}
\put(30,30){\vector(0,1){20}}
\put(25,50){$x$}
\put(50,25){$t$}

\put(220,-40){\makebox(0,0)[b]{\bf {Figure 1}}}
\end{picture}
\end{center}

\vspace{2cm}

The curve for this theory is given by eq. (\ref{infcurve}),
with $\arh=1$. 
A particular parametrization for the functions 
$P_n(x)$ and $J_n(x)$  was given in ref.~\cite{twoanti}.
In this paper, as discussed above, 
we choose a different parametrization,
one in which the $P_n(x)$ and $J_n(x)$ have the symmetries
\be
P_{n + 2\ell }(x)&=&P_n (x-\ell\Delta),\nn\\
J_{n + 2\ell }(x)&=&J_n (x-\ell\Delta),
\label{twobranesagain}
\ee
where $m = \ha (m_1 + m_2)$ and $\Delta=m_1-m_2$. 
In this parametrization, the curve becomes
\be
\sum_{n\,{\rm  even}}\, L^{n^2} t^n 
\, J_0(x-\nha n\Delta) P_0(x-\nha n\Delta)\,+
\sum_{\rm n \,{\rm odd}}\, L^{n^2 } t^n 
\, J_1(x-\nha (n-1) \Delta)\, \ P_1(x-\nha (n-1) \Delta)\,=\,0,\nn\\
\label{twoanticurve}
\ee
with
\be
J_{0}(x)=
\prod_{j=1}^{N_f}
\prod_{p=1}^{\infty} \,
\left( [x+\nha m + (-1)^p (M_j -\nha m) - \nha p \Delta]
       [x+\nha m + (-1)^p (M_j -\nha m) + \nha p \Delta] \right)^{p/2},\nn\\
\ee
and with the leading term (in $L$) of $P_0(x)$ given by
\be
P_{0}(x)&=&
\prod_{p=1}^{\infty} \,
\left([x+\nha m+\nha (p-\ha)\Delta] [x+\nha m-\nha (p-\ha)\Delta] \right)^{-2p}
\prod_{i=1}^N (x-a_i).
\label{figunoone}
\ee
Using the involution symmetry in the O$6^-$ plane at $x = -\nha m_2$,
one obtains
\be
P_{1}(x)&=&P_{0}(-x-\xxt m+ \nha \Delta),\nn\\
J_{1}(x)&=&J_{0}(-x-\xxt m+ \nha \Delta).
\label{odd}
\ee

For $N_f=4$, the beta function vanishes,
so the curve for this model is (\ref{ellipcurve}), with $r=1$.
When the masses of the hypermultiplets satisfy
$m_1 = m_2 = 2 M_j$~$(j=1,\ldots,4)$, 
the curve becomes
\be
0 &=&
\sum_{n\,{\rm even}}\, q^{n^2/4} \, e^{\beta n z} \,\prod_{i=1}^{N}(x-a_i)\,+
\sum_{n\,{\rm odd}} \, q^{n^2/4} \, e^{\beta n z} 
\,(-1)^N\prod_{i=1}^{N}(x+a_i+\xxt m)   \nn \\
&=&
\theta_3 \left({z \over  \omega_1} | 2\tau\right) \, \prod_{i=1}^{N}(x-a_i)\,+
\theta_2 \left({z \over  \omega_1} | 2\tau\right)
 \,(-1)^N\prod_{i=1}^{N}(x+a_i+ \xxt m),
\label{specialcase}
\ee
with no subleading terms,
where $\theta_2$ and  $\theta_3$ are Jacobi theta functions (\ref{thetafcns}). 
The curve (\ref{specialcase}) 
is manifestly invariant under $z \to z+2\omega_2$.

Using eq.~(\ref{masterfcn}), we find
\be
S(x)
= {\xf \prod_{i=1}^{N}(x+a_i+\xxt m_1) \prod_{i=1}^N(x +a_i+\xxt m_2) 
\prod_{j=1}^{N_f}(x +M_j) \over (x +\nha m_1)^2 (x +\nha m_2)^2
\prod_{i=1}^N (x-a_i)^2}, ~~~~~~~~~
\label{ffuno}
\ee
in agreement with the empirical rules given in Table 3.
The form of $S(x)$ is independent of the parametrization used
for $P_n(x)$ and $J_n(x)$.
The one-instanton contribution to the prepotential is \cite{twoanti}
\be 
\xe \pi i {\cal F}_{\rm 1-inst}=
\sum_{k=1}^N S_k(a_k) -2 S_{m_1}(-\nha m_1)-2S_{m_2}(-\nha m_2),
\ee
where 
\be
S_k(a_k)
&=& {\xf \prod_{i=1}^{N}(a_k+a_i+\xxt m_1) \prod_{i=1}^N(a_k+a_i+\xxt m_2) 
\prod_{j=1}^{N_f}(a_k +M_j) \over (a_k +\nha m_1)^2 (a_k +\nha m_2)^2
\prod_{i\neq k}^N (a_k-a_i)^2},  \nn\\ \nn\\
S_{m_1}(-\nha m_1)
&=& {\xf \prod_{i=1}^N(a_i+\xxt m_2 - \nha m_1) \prod_{j=1}^{N_f}(M_j-\nha m_1) 
\over  (\nha m_2 -\nha m_1)^2 \prod_{i=1}^N (a_i + \nha m_1)}, \nn\\ \nn\\
S_{m_2}(-\nha m_2)
&=& {\xf \prod_{i=1}^N(a_i+\xxt m_1 - \nha m_2) \prod_{j=1}^{N_f}(M_j-\nha m_2) 
\over  (\nha m_1 -\nha m_2)^2 \prod_{i=1}^N (a_i + \nha m_2)}.
\ee
Various checks of this result were made in ref. \cite{twoanti}.
(In ref.~\cite{twoanti}, the masses of the antisymmetric hypermultiplets
were defined to be $2m_1$ and $2m_2$ rather than $m_1$ and $m_2$,
and the definition of $S(x)$ differed by a factor of 4.)

\eject

\noindent{\bf 4.2.} 
\underline{SU$(N)\,+\, 1\,{\rm antisymmetric} + 1\,{\rm symmetric}$}
\renewcommand{\theequation}{4.2.\arabic{equation}}
\setcounter{equation}{0}

Consider SU$(N)$ gauge theory 
with one matter hypermultiplet (mass $m_1$) in the antisymmetric representation,
and one (mass $m_2$) in the symmetric representation.
The corresponding M-theory picture, represented in Fig. 2,
contains O$6^+$ planes (related to the symmetric hypermultiplet)
coincident with the even NS 5-branes,
and O$6^-$ planes (related to the antisymmetric hypermultiplet)
coincident with the odd NS 5-branes.

\begin{center}
\begin{picture}(810,295)(10,10)


\put(10,192){\line(1,0){5}}
\put(20,192){\line(1,0){5}}
\put(30,192){\line(1,0){5}}
\put(40,192){\line(1,0){5}}
\put(50,192){\line(1,0){5}}
\put(60,192){\line(1,0){5}}
\put(70,192){\line(1,0){5}}
\put(80,192){\line(1,0){5}}
\put(90,192){\line(1,0){5}}
\put(100,192){\line(1,0){5}}
\put(5,180){$(\!x\!-\!a_i\!+\!\xxt m_2\!-\!\xxt m_1\!)$}

\put(105,240){\line(1,0){2}}
\put(110,240){\line(1,0){5}}
\put(120,240){\line(1,0){5}}
\put(130,240){\line(1,0){5}}
\put(140,240){\line(1,0){5}}
\put(150,240){\line(1,0){5}}
\put(160,240){\line(1,0){5}}
\put(170,240){\line(1,0){5}}
\put(180,240){\line(1,0){5}}
\put(190,240){\line(1,0){5}}
\put(200,240){\line(1,0){5}}
\put(120,245){$(\!x\!+\!a_i\!+\!\xxt m_2\!)$}

\put(50,-10){$P_{2}$}
\put(150,-10){$P_{1}$}
\put(250,-10){$P_0$}
\put(350,-10){$P_{-1}$}
\put(450,-10){$P_{-2}$}

\put(205,117){\line(1,0){2}}
\put(210,117){\line(1,0){5}}
\put(220,117){\line(1,0){5}}
\put(230,117){\line(1,0){5}}
\put(240,117){\line(1,0){5}}
\put(250,117){\line(1,0){5}}
\put(260,117){\line(1,0){5}}
\put(270,117){\line(1,0){5}}
\put(280,117){\line(1,0){5}}
\put(290,117){\line(1,0){5}}
\put(300,117){\line(1,0){5}}
\put(235,105){$(\!x-\!a_i\!)$}

\put(305,155){\line(1,0){2}}
\put(310,155){\line(1,0){5}}
\put(320,155){\line(1,0){5}}
\put(330,155){\line(1,0){5}}
\put(340,155){\line(1,0){5}}
\put(350,155){\line(1,0){5}}
\put(360,155){\line(1,0){5}}
\put(370,155){\line(1,0){5}}
\put(380,155){\line(1,0){5}}
\put(390,155){\line(1,0){5}}
\put(400,155){\line(1,0){5}}
\put(320,160){$(\!x\!+\!a_i\!+\!\xxt m_1\!)$}

\put(405,10){\line(1,0){2}}
\put(410,10){\line(1,0){5}}
\put(420,10){\line(1,0){5}}
\put(430,10){\line(1,0){5}}
\put(440,10){\line(1,0){5}}
\put(450,10){\line(1,0){5}}
\put(460,10){\line(1,0){5}}
\put(470,10){\line(1,0){5}}
\put(480,10){\line(1,0){5}}
\put(490,10){\line(1,0){5}}
\put(500,10){\line(1,0){5}}
\put(407,15){$(\!x\!-\!a_i\!+\!\xxt m_1\!-\!\xxt m_2\!)$}


\put(105,5){\line(0,1){255}}
\put(205,5){\line(0,1){255}}
\put(305,5){\line(0,1){255}}
\put(405,5){\line(0,1){255}}
\put(103,271){$\NSone$}
\put(203,271){$\NStwo$}
\put(303,271){$\NSthree$}
\put(403,271){$\NSfour$}


\put(101,215){$\otimes$} 
\put(201,175){$\otimes$}
\put(301,135){$\otimes$}
\put(401,85){$\otimes$}
\put(31,220){$(\!x\!+\!\xxt m_2\!-\!\nha m_1\!)$}
\put(160,170){$(\!x\!+\!\nha m_2\!)$}
\put(259,135){$(\!x\!+\!\nha m_1\!)$}
\put(406,75){$(\!x\!+\!\xxt m_1\!-\!\nha m_2\!)$}

\put(110,220){O$6^{-}$}
\put(210,180){O$6^{+}$}
\put(310,140){O$6^{-}$}
\put(410,90){O$6^{+}$}

\put(30,30){\vector(1,0){20}}
\put(30,30){\vector(0,1){20}}
\put(25,50){$x$}
\put(50,25){$t$}

\put(220,-40){\makebox(0,0)[b]{\bf {Figure 2}}}
\end{picture}
\end{center}

\vspace{4cm}

Choosing the parametrization of the coefficient functions of
the curve to have the properties (\ref{twobranes}),
the curve (\ref{ellipcurve}) for this theory takes the form
\be
\sum_{n\,{\rm  even}}\, q^{n^2 / 4} e^{\beta n z}
\, P_0(x-\nha n\Delta)\,+
\sum_{n\,{\rm odd}}\, q^{n^2 / 4} e^{\beta n z}
\ P_1(x-\nha (n-1) \Delta)\,=\,0,
\label{zfour}
\ee
where the leading term of $P_0(x)$ is 
\be
P_{0}(x)
&=& 
\prod_{p=1}^{\infty} 
\left[ x+ \nha m+ \nha (-1)^p (p-\ha)\Delta 
\over  x+\nha m-\nha (-1)^p (p-\ha)\Delta \right]^{2p}
\prod_{i=1}^N (x-a_i),
\label{zone}
\ee
and, using the involution property,
\be
P_{1}(x)\,=\,P_{0}(-x-\xxt m+\nha \Delta),
\label{ztwo}
\ee
with $m = \ha (m_1 + m_2)$ and $\Delta=m_1-m_2$. 
The curve (\ref{zfour}) is manifestly invariant under
$z \to z+2\omega_2$ together with $x \to x+\Delta$.

It may be verified that in the 
$m_2\to \infty$ ($m_1\to \infty$) limit, 
the curve (\ref{zfour}) reduces 
to the curve (leading-order coefficients only) 
for SU$(N)$ with one antisymmetric (symmetric)
hypermultiplet \cite{LandsteinerLopezLowe}.

In the case $m_1=m_2$,
{\ie}, zero global mass $\Delta$, 
the subleading terms of $P_n(x)$ vanish
(the effects of the orientifolds of opposite charge cancel, 
as they are located at the same position in the $x$ plane), 
and the curve (\ref{zfour}) 
reduces to eq.~(\ref{specialcase}),
which, surprisingly, also describes
SU$(N)$ with two antisymmetric hypermultiplets (masses $m_1$ and $m_2$) 
and four fundamental hypermultiplets (masses $M_j$), 
with $m_1=m_2=2M_j$.
(See section 5 for more details.)

Using eq.~(\ref{masterfcn}), we obtain
\be
S(x) \,=\,
{\xf  (x+\nha m_2)^2\,
\prod_{i=1}^N (x+a_i+\xxt m_1)  \prod_{i=1}^N (x+a_i+\xxt m_2) \over
(x+\nha m_1)^2 \prod_{i=1}^N (x-a_i)^2}.
\label{defino}
\ee
The one-instanton contribution to the prepotential is given by 
\be 
\xe \pi i {\cal F}_{\rm 1-inst}=
\sum_{k=1}^{N} S_k(a_k)-2S_{m_1}(-\nha m_1),
\label{xxdosthree} 
\ee 
with 
\be
S_k(a_k)&=&
{\xf \,(a_k+\nha m_2)^2\, 
\prod_{i=1}^{N}  (a_k+a_i+\xxt m_1) \,\prod_{i=1}^N(a_k +a_i+\xxt m_2) 
\over (a_k +\nha m_1)^2 \prod_{i\neq k}^N (a_k-a_i)^2};\nn\\ \nn\\
S_{m_1}(-\nha m_1)
&=&
{(m_1-m_2)^2\,\prod_{i=1}^N(a_i+\xxt m_2 - \nha m_1) \over
4\,\prod_{i=1}^N (a_i+\nha m_1)}.
\label{xxuno}
\ee

We can test the expression (\ref{xxdosthree}) 
in two particular cases. 
For SU(2), 
we can compare our results against those for SU(2) + adjoint, 
as given in eqs.~(\ref{duno}) and (\ref{dunotwo}). 
Setting  $\xxt m_2=m$ in (\ref{xxuno}), 
with $m$ the mass of the adjoint hypermultiplet, and $a_1+a_2 = 0$, 
we find that both expressions agree up to a moduli 
independent additive constant.
For SU(3),
we can compare the prepotential (\ref{xxdosthree})
against that for SU(3) with one symmetric representation and 
one fundamental representation \cite{nonhyper},
finding agreement (after setting $a_1+a_2+a_3=0$ and $m_1 = - m_f$) 
up to a moduli independent additive constant.

\eject

\noindent{\bf 4.3.} 
\underline{SU$(N)\,+\,{\rm adjoint}$}
\renewcommand{\theequation}{4.3.\arabic{equation}}
\setcounter{equation}{0}

Consider SU$(N)$ gauge theory 
with  a matter hypermultiplet (mass $m$) in the adjoint representation.
This is an elliptic model which can be described in terms of the 
M-theory  picture in  Fig. 3.a.
In this figure,  there are $N$ D4-branes suspended between a single 
NS 5-brane with a periodicity in $t$, 
but with a shift in $x$ of $m$ (the global mass) for each circuit of $t$. 
The covering space of the $S^1$ (the $t$-variable) is shown in Fig. 3.b.

\begin{figure} [h]
\centerline{ 
\psfig{figure=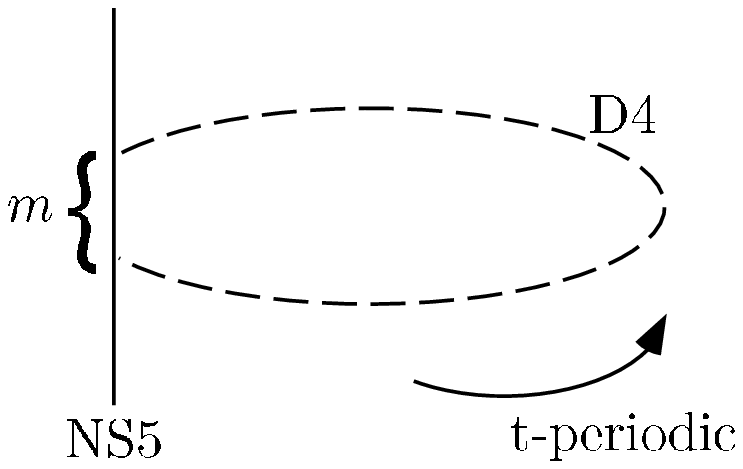,height=3.6cm,width=4.6cm}}
\begin{center}
{\footnotesize{\bf Figure 3.a}}
\end{center}
\end{figure}

\begin{picture}(600,140)(10,10)

\put(40,40){\line(0,1){100}}
\put(130,40){\line(0,1){100}}
\put(220,40){\line(0,1){100}}
\put(310,40){\line(0,1){100}}
\put(400,40){\line(0,1){100}}
\put(-10,50){\line(1,0){10}}
\put(10,50){\line(1,0){10}}
\put(30,50){\line(1,0){10}}
\put(-20,55){$(\!x\!-\!a_i\!-\!2m\!)$}
\put(10,20){$P_2$}

\put(40,65){\line(1,0){10}}
\put(60,65){\line(1,0){10}}
\put(80,65){\line(1,0){10}}
\put(100,65){\line(1,0){10}}
\put(120,65){\line(1,0){10}}
\put(60,75){$(\!x\!-\!a_i\!-\!m\!)$}
\put(80,20){$P_1$}

\put(130,80){\line(1,0){10}}
\put(150,80){\line(1,0){10}}
\put(170,80){\line(1,0){10}}
\put(190,80){\line(1,0){10}}
\put(210,80){\line(1,0){10}}
\put(155,85){$(\!x\!-\!a_i\!)$}
\put(165,20){$P_0$}

\put(220,95){\line(1,0){10}}
\put(240,95){\line(1,0){10}}
\put(260,95){\line(1,0){10}}
\put(280,95){\line(1,0){10}}
\put(300,95){\line(1,0){10}}
\put(240,100){$(\!x\!-\!a_i\!+\!m\!)$}
\put(250,20){$P_{-1}$}

\put(310,110){\line(1,0){10}}
\put(330,110){\line(1,0){10}}
\put(350,110){\line(1,0){10}}
\put(370,110){\line(1,0){10}}
\put(390,110){\line(1,0){10}}
\put(320,115){$(\!x\!-\!a_i\!+\!2m\!)$}
\put(325,20){$P_{-2}$}

\put(400,125){\line(1,0){10}}
\put(420,125){\line(1,0){10}}
\put(440,125){\line(1,0){10}}
\put(410,130){$(\!x\!-\!a_i\!+\!3m\!)$}
\put(425,20){$P_{-3}$}

\put(100,0){\vector(1,0){35}}
\put(140,0){${\it t\ (\rm{covering\ of}\ S^1)}$}

\put(10,100){\vector(0,1){30}}
\put(15,110){$x$}
\put(220,-20){\makebox(0,0)[b]{\bf {Figure 3.b}}}

\end{picture}
\vspace{.2in}

The coefficient functions (which have no subleading terms) are
\be
P_0(x)&=&\prod_{i=1}^N (x-a_i)\,,\nn\\
P_n(x)&=& P_0(x-nm),
\label{zten}
\ee
so the curve (\ref{ellipcurve}), with $\arh=2$, becomes
\be
\sum_n  q^{n^2/2} e^{\beta n z} \prod_{i=1}^N (x-a_i-nm)\,=\,0.
\label{sunadjoint}
\ee
This is manifestly invariant under
$z \to z+2\omega_2$ together with $x \to x+m$.
Shifting $ z \to z + \omega_1 + \omega_2$,
eq.~(\ref{sunadjoint})
becomes identical to the result of D'Hoker and Phong \cite{DHokerPhong}, 
\be
\sum_n (-1)^n q^{n(n-1)/ 2} e^{\beta n z} \prod_{i=1}^N (x-a_i-nm)\,=\,0.
\label{zeleven}
\ee
In section 6, we show its relation to the curve 
of Donagi and Witten \cite{DonagiWitten} for this theory.

Using eq.~(\ref{masterfcn}), we obtain
\be
S(x)\,=\,{\xf \prod_{i=1}^N[(x-a_i)^2-m^2] \over \prod_{i=1}^N(x-a_i)^2}.
\label{chorizo}
\ee
The one-instanton contribution to the prepotential 
is given by \cite{DHokerPhong}
\be
\xe \pi i {\cal F}_{\rm 1-inst}\,=\, \sum_{k=1}^N S_k(a_k), 
\label{duno}
\ee
where
\be
S_k(a_k)\,=\,{\xf \prod_{i=1}^N [(a_k-a_i)^2-m^2] 
\over {\prod_{i\neq k}^N (a_k-a_i)^2}}.
\label{dunotwo}
\ee
No subtraction is required in eq. (\ref{duno}), 
as there are no spurious singularities to remove from $\sum_k\, S_k(a_k)$. 
This fact is related to the absence of subleading terms.
\eject

\noindent{\bf 4.4.} 
\underline{SO$(2N)\,+\, {\rm adjoint}$}
\renewcommand{\theequation}{4.4.\arabic{equation}}
\setcounter{equation}{0}

Consider SO$(2N)$ gauge theory 
with a matter hypermultiplet (mass $m$) in the adjoint representation.
The corresponding M-theory picture (Fig. 4) 
contains O$6^-$ planes on top of each NS 5-brane, 
O$6^+$ planes between each pair of NS 5-branes, 
and D4-branes. 
There are additional fourbranes in mirror positions 
with respect to the O$6^+$ orientifolds 
that are {\it not} represented in the figure.

\begin{center}
\begin{picture}(810,295)(10,10)


\put(10,192){\line(1,0){5}}
\put(20,192){\line(1,0){5}}
\put(30,192){\line(1,0){5}}
\put(40,192){\line(1,0){5}}
\put(50,192){\line(1,0){5}}
\put(60,192){\line(1,0){5}}
\put(70,192){\line(1,0){5}}
\put(80,192){\line(1,0){5}}
\put(90,192){\line(1,0){5}}
\put(100,192){\line(1,0){5}}
\put(10,180){$(\!x\!-\!2m\!-\!a_i\!)$}

\put(105,250){\line(1,0){2}}
\put(110,250){\line(1,0){5}}
\put(120,250){\line(1,0){5}}
\put(130,250){\line(1,0){5}}
\put(140,250){\line(1,0){5}}
\put(150,250){\line(1,0){5}}
\put(160,250){\line(1,0){5}}
\put(170,250){\line(1,0){5}}
\put(180,250){\line(1,0){5}}
\put(190,250){\line(1,0){5}}
\put(200,250){\line(1,0){5}}
\put(125,255){$(\!x\!-\!m\!+\!a_i\!)$}

\put(205,117){\line(1,0){2}}
\put(210,117){\line(1,0){5}}
\put(220,117){\line(1,0){5}}
\put(230,117){\line(1,0){5}}
\put(240,117){\line(1,0){5}}
\put(250,117){\line(1,0){5}}
\put(260,117){\line(1,0){5}}
\put(270,117){\line(1,0){5}}
\put(280,117){\line(1,0){5}}
\put(290,117){\line(1,0){5}}
\put(300,117){\line(1,0){5}}
\put(235,105){$(\!x\!-\!a_i\!)$}

\put(305,155){\line(1,0){2}}
\put(310,155){\line(1,0){5}}
\put(320,155){\line(1,0){5}}
\put(330,155){\line(1,0){5}}
\put(340,155){\line(1,0){5}}
\put(350,155){\line(1,0){5}}
\put(360,155){\line(1,0){5}}
\put(370,155){\line(1,0){5}}
\put(380,155){\line(1,0){5}}
\put(390,155){\line(1,0){5}}
\put(400,155){\line(1,0){5}}
\put(320,160){$(\!x\!+\!m\!+\!a_i\!)$}

\put(405,10){\line(1,0){2}}
\put(410,10){\line(1,0){5}}
\put(420,10){\line(1,0){5}}
\put(430,10){\line(1,0){5}}
\put(440,10){\line(1,0){5}}
\put(450,10){\line(1,0){5}}
\put(460,10){\line(1,0){5}}
\put(470,10){\line(1,0){5}}
\put(480,10){\line(1,0){5}}
\put(490,10){\line(1,0){5}}
\put(500,10){\line(1,0){5}}
\put(415,15){$(\!x\!+\!2m\!-\!a_i\!)$}


\put(105,5){\line(0,1){255}}
\put(205,5){\line(0,1){255}}
\put(305,5){\line(0,1){255}}
\put(405,5){\line(0,1){255}}
\put(103,271){$\NSone$}
\put(203,271){$\NStwo$}
\put(303,271){$\NSthree$}
\put(403,271){$\NSfour$}


\put(51,235){$\otimes$} 
\put(51,245){O$6^{+}$}
\put(10,235){$(\!x\!-\!2m\!)$}
\put(101,215){$\otimes$} 
\put(110,215){O$6^{-}$}
\put(201,175){$\otimes$}
\put(210,175){O$6^{-}$}
\put(301,135){$\otimes$}
\put(310,135){O$6^{-}$}
\put(401,95){$\otimes$}
\put(410,95){O$6^{-}$}
\put(60,215){$(\!x\!-\!\thh m\!)$}
\put(160,168){$(\!x\!-\!\ha m\!)$}
\put(260,130){$(\!x\!+\!\ha m\!)$}

\put(151,195){$\otimes$}
\put(151,205){O$6^{+}$}
\put(251,155){$\otimes$}
\put(251,165){O$6^{+}$}
\put(351,115){$\otimes$}
\put(351,125){O$6^{+}$}
\put(451,75){$\otimes$}
\put(451,85){O$6^{+}$}
\put(430,60){$(\!x\!+\!2m\!)$}

\put(130,184){$(\!x\!-\!m\!)$}
\put(362,85){$(\!x\!+\!\thh m\!)$}
\put(248,148){$\!x\!$}
\put(340,105){$(\!x+\!m\!)$}

\put(30,30){\vector(1,0){20}}
\put(30,30){\vector(0,1){20}}
\put(25,50){$x$}
\put(50,25){$t$}

\put(50,-10){$P_{2}$}
\put(150,-10){$P_{1}$}
\put(250,-10){$P_0$}
\put(350,-10){$P_{-1}$}
\put(450,-10){$P_{-2}$}

\put(220,-30){\makebox(0,0)[b]{\bf {Figure 4}}}
\end{picture}
\end{center}

\vspace{1cm}

In this theory, 
the unit cell contains only one NS 5-brane,
so the parametrization of the coefficient functions 
of the curve are chosen to respect
\be
P_{n + \ell}(x) = P_n (x-\ell m).
\label{onebraneagain}
\ee
With this, the curve (\ref{ellipcurve}) becomes
\be
\sum_{n=-\infty}^\infty q^{n^2 / 2} e^{\beta n z} P_0(x-nm)\,=\,0, 
\label{zfourteen}
\ee
where the leading term  of $P_0(x)$ is given by 
\be
P_0(x)\,=\,\prod_{p=1}^{\infty} 
\left[(x+pm)(x-pm) \over (x+(p-\ha)m ) (x-(p-\ha)m) \right]^{2p}
\prod_{i=1}^N [x^2-a_i^2].
\label{ztwelve}
\ee
The curve (\ref{zfourteen}) is manifestly invariant under
$z \to z+2\omega_2$ and $x \to x+m$.

There are several decoupling limits \cite{Yokono} 
that can be considered  to
check eq.~(\ref{zfourteen}). 

\noindent
(i) Let
\be
x_0\,=\,{1 \over N} \sum_{i=1}^N a_i,
\label{zazfourteen}
\ee
be the average position of the D4-branes in a single
cell which lie above the O$6^{+}$ inside that cell. 
Then change variables
\be
x \to x + x_0,\,\,\,\,\,\,\,\,\, a_i \to a_i + x_0, 
\label{zzfourteen}
\ee
and let $x_0 \rightarrow \infty$. 
After taking  these limits in eq. (\ref{zfourteen}), the final 
curve coincides  with the one for ${\cal N}=4$  SU$(N)$ + adjoint. 

\noindent
(ii) There is another decoupling limit we can consider. If we take 
\be
&& q \to 0\,\,\,(\tau \to i\infty)\,\,,\,\,\, \,\,\,\,
m \to \infty \,\,\,\,,\,\,\Lambda^{b_0} =q m^{b_0}\,\,\,\,{\rm fixed}; \nn \\
&& b_0\,=\,4N_1-2N_2-4\,\,,\,\,\,\,\,\,\,N_1+N_2\,=\,N,
\label{decoupling}
\ee
we obtain the curve corresponding to 
${\cal N}=2$ SO$(2N_1)$ + $N_2$ fundamentals. 

Using eq. (\ref{masterfcn}), we obtain
\be 
S(x)\,=\,{\xf x^4
\prod_{i=1}^N[(x-m)^2-a_i^2]\prod_{i=1}^N[(x+m)^2-a_i^2]\over 
{(x+\ha m)^2(x-\ha m)^2
\prod_{i=1}^N (x^2-a_i^2)^2}}. 
\label{tresone} 
\ee 
The one-instanton contribution to the prepotential is given by
\be 
\xe \pi i
{\cal F}_{\rm 1-inst}\,=\, \sum_{k=1}^N S_k(a_k)-2 S_m(-\ha m),
\label{trestwo} 
\ee 
with
\be
S_k(a_k)&=&{\xf  a_k^2\,
\prod_{i=1}^N [(a_k-m)^2 -a_i^2]\, \prod_{i=1}^N [(a_k+m)^2 -a_i^2]
\over { 4 (a_k+\ha m)^2 (a_k-\ha m)^2 \,\prod_{i\neq k}^N (a_k^2-a_i^2)^2}},
\nn\\ \nn\\
S_m(-\nha m)&=&{\xf \,m^2\,
\prod_{i=1}^N ({9 \over 4} m^2 -a_i^2)
\over { 16  \prod_{i=1}^N  ( {1\over 4} m^2 -a_i^2)}}.
\label{zzsixteen}
\ee

We can  compare our result for 
SO$(2N)$ with a massive adjoint hypermultiplet
with that of Minahan et. al. \cite{Minahan}, 
who obtained a mass expansion for ${\cal F}_{\rm instanton}$,
using a conjectured recursion relation for SO$(2N)$ plus adjoint. 
Expanding eqs.~(\ref{trestwo})--(\ref{zzsixteen}) 
in powers of $m$, we find the first moduli-dependent contribution to be 
\be
\xe \pi i {\cal F}_{\rm 1-inst}\,
=\,  4 m^4  \sum_{k=1}^N \sum_{j\neq k} 
{a_k^2\over (a_k^2-a_j^2)^2}\,+\,{\cal O}(m^6). 
\label{tresthree}
\ee 
This coincides with eq. (4.7) of ref.~\cite{Minahan},  
identifying their $\phi_k$ with our $a_k$.

We can make a further check of our proposed expression (\ref{trestwo})
by comparing the result for SO(6) with that for SU(4).
The one-instanton prepotential for  SU$(4)$ with a massive adjoint
hypermultiplet is given by eq.~(\ref{duno}),
with moduli $a'_k$, $k=1,\ldots, 4$,
restricted by $\sum_{k=1}^4 a'_k=0$. 
This constraint allows us to eliminate $a'_4$.
Using the change of variables 
\be
a_1= a'_1+a'_2,\qquad a_2=a'_2+a'_3,\qquad a_3=a'_1+a'_3,
\label{cambio}
\ee
where $a_i$ are the moduli of SO(6),
we find that  the one-instanton prepotentials
(\ref{duno}) and (\ref{trestwo}) agree,
up to a moduli-independent additive constant.

To construct our curve, we have used the M-theory picture 
suggested by Uranga \cite{Uranga} in terms of O$6^{\pm}$ planes. 
Yokono \cite{Yokono} has constructed curves for SO$(2N)$ 
using orientifold fourplanes instead of O$6^{\pm}$ planes.
His curves, while satisfying the correct decoupling limits (i) and (ii) above,
differ from ours.
Although the curves in ref. \cite{Yokono} have a smooth limit when $m \to 0$, 
the brane configuration from which it is 
constructed  is not consistent in this limit. 
When $m=0$, the fourbranes 
must change their charge when crossing an  NS 5-brane \cite{Evans}, 
but this is not the case for the M-theory picture in ref. \cite{Yokono}. 
(The same comments apply to the curves 
proposed for Sp$(2N)$ + adjoint and SO$(2N+1)$ + adjoint in ref.~\cite{Yokono}.)
Moreover, the one-loop prepotential derived from Yokono's curves
disagrees with the perturbation theory result, 
and the one-instanton prepotential for SO(6) with massive adjoint 
calculated from his curve conflicts with that for SU(4) with massive adjoint.
\eject

\noindent{\bf 4.5.} 
\underline{SO$(2N+1)\,+\, {\rm adjoint}$}
\renewcommand{\theequation}{4.5.\arabic{equation}}
\setcounter{equation}{0}

Consider SO$(2N+1)$ gauge theory 
with a matter hypermultiplet (mass $m$) in the adjoint representation.
The M-theory picture in Fig. 5 is similar to the one for 
SO$(2N)$ with adjoint hypermultiplet with the following difference:
there is an additional fourbrane whose position is fixed at the O$6^{+}$ plane
in each cell \cite{Uranga}. 
As in last sections, 
there are additional fourbranes at mirror positions that are
not included in the figure.

\begin{center}
\begin{picture}(810,295)(10,10)


\put(10,192){\line(1,0){5}}
\put(20,192){\line(1,0){5}}
\put(30,192){\line(1,0){5}}
\put(40,192){\line(1,0){5}}
\put(50,192){\line(1,0){5}}
\put(60,192){\line(1,0){5}}
\put(70,192){\line(1,0){5}}
\put(80,192){\line(1,0){5}}
\put(90,192){\line(1,0){5}}
\put(100,192){\line(1,0){5}}
\put(10,182){$(\!x\!-\!2m\!-\!a_i\!)$}

\put(105,250){\line(1,0){2}}
\put(110,250){\line(1,0){5}}
\put(120,250){\line(1,0){5}}
\put(130,250){\line(1,0){5}}
\put(140,250){\line(1,0){5}}
\put(150,250){\line(1,0){5}}
\put(160,250){\line(1,0){5}}
\put(170,250){\line(1,0){5}}
\put(180,250){\line(1,0){5}}
\put(190,250){\line(1,0){5}}
\put(200,250){\line(1,0){5}}
\put(125,255){$(\!x\!-\!m\!+\!a_i\!)$}

\put(205,100){\line(1,0){2}}
\put(210,100){\line(1,0){5}}
\put(220,100){\line(1,0){5}}
\put(230,100){\line(1,0){5}}
\put(240,100){\line(1,0){5}}
\put(250,100){\line(1,0){5}}
\put(260,100){\line(1,0){5}}
\put(270,100){\line(1,0){5}}
\put(280,100){\line(1,0){5}}
\put(290,100){\line(1,0){5}}
\put(300,100){\line(1,0){5}}
\put(235,88){$(\!x\!-\!a_i\!)$}

\put(305,170){\line(1,0){2}}
\put(310,170){\line(1,0){5}}
\put(320,170){\line(1,0){5}}
\put(330,170){\line(1,0){5}}
\put(340,170){\line(1,0){5}}
\put(350,170){\line(1,0){5}}
\put(360,170){\line(1,0){5}}
\put(370,170){\line(1,0){5}}
\put(380,170){\line(1,0){5}}
\put(390,170){\line(1,0){5}}
\put(400,170){\line(1,0){5}}
\put(320,175){$(\!x\!+\!m\!+\!a_i\!)$}

\put(405,10){\line(1,0){2}}
\put(410,10){\line(1,0){5}}
\put(420,10){\line(1,0){5}}
\put(430,10){\line(1,0){5}}
\put(440,10){\line(1,0){5}}
\put(450,10){\line(1,0){5}}
\put(460,10){\line(1,0){5}}
\put(470,10){\line(1,0){5}}
\put(480,10){\line(1,0){5}}
\put(490,10){\line(1,0){5}}
\put(500,10){\line(1,0){5}}
\put(415,15){$(\!x\!+\!2m\!-\!a_i\!)$}


\put(105,5){\line(0,1){255}}
\put(205,5){\line(0,1){255}}
\put(305,5){\line(0,1){255}}
\put(405,5){\line(0,1){255}}
\put(103,271){$\NSone$}
\put(203,271){$\NStwo$}
\put(303,271){$\NSthree$}
\put(403,271){$\NSfour$}


\put(51,235){$\otimes$} 
\put(51,245){O$6^{+}$}
\put(10,237){\line(1,0){5}}
\put(20,237){\line(1,0){5}}
\put(30,237){\line(1,0){5}}
\put(40,237){\line(1,0){5}}
\put(60,237){\line(1,0){5}}
\put(70,237){\line(1,0){5}}
\put(80,237){\line(1,0){5}}
\put(90,237){\line(1,0){5}}
\put(100,237){\line(1,0){5}}

\put(10,245){$(\!x\!-\!2m\!)$}
\put(101,215){$\otimes$} 
\put(110,215){O$6^{-}$}
\put(201,175){$\otimes$}
\put(210,175){O$6^{-}$}
\put(301,135){$\otimes$}
\put(310,135){O$6^{-}$}
\put(401,95){$\otimes$}
\put(410,95){O$6^{-}$}
\put(62,215){$(\!x\!-\!\thh m\!)$}
\put(160,168){$(\!x\!-\!\ha m\!)$}
\put(260,130){$(\!x\!+\!\ha m\!)$}

\put(151,195){$\otimes$}
\put(151,205){O$6^{+}$}
\put(105,197){\line(1,0){2}}
\put(110,197){\line(1,0){5}}
\put(120,197){\line(1,0){5}}
\put(130,197){\line(1,0){5}}
\put(140,197){\line(1,0){5}}
\put(160,197){\line(1,0){5}}
\put(170,197){\line(1,0){5}}
\put(180,197){\line(1,0){5}}
\put(190,197){\line(1,0){5}}
\put(200,197){\line(1,0){5}}

\put(251,155){$\otimes$}
\put(251,165){O$6^{+}$}
\put(205,157){\line(1,0){2}}
\put(210,157){\line(1,0){5}}
\put(220,157){\line(1,0){5}}
\put(230,157){\line(1,0){5}}
\put(240,157){\line(1,0){5}}
\put(260,157){\line(1,0){5}}
\put(270,157){\line(1,0){5}}
\put(280,157){\line(1,0){5}}
\put(290,157){\line(1,0){5}}
\put(300,157){\line(1,0){5}}

\put(351,115){$\otimes$}
\put(351,125){O$6^{+}$}
\put(305,117){\line(1,0){2}}
\put(310,117){\line(1,0){5}}
\put(320,117){\line(1,0){5}}
\put(330,117){\line(1,0){5}}
\put(340,117){\line(1,0){5}}
\put(360,117){\line(1,0){5}}
\put(370,117){\line(1,0){5}}
\put(380,117){\line(1,0){5}}
\put(390,117){\line(1,0){5}}
\put(400,117){\line(1,0){5}}

\put(451,75){$\otimes$}
\put(451,85){O$6^{+}$}
\put(405,77){\line(1,0){2}}
\put(410,77){\line(1,0){5}}
\put(420,77){\line(1,0){5}}
\put(430,77){\line(1,0){5}}
\put(440,77){\line(1,0){5}}
\put(460,77){\line(1,0){5}}
\put(470,77){\line(1,0){5}}
\put(480,77){\line(1,0){5}}
\put(490,77){\line(1,0){5}}
\put(500,77){\line(1,0){5}}

\put(430,60){$(\!x\!+\!2m\!)$}

\put(130,184){$(\!x\!-\!m\!)$}
\put(362,85){$(\!x\!+\!\thh m\!)$}
\put(248,148){$\!x\!$}
\put(340,105){$(\!x+\!m\!)$}

\put(30,30){\vector(1,0){20}}
\put(30,30){\vector(0,1){20}}
\put(25,50){$x$}
\put(50,25){$t$}

\put(50,-10){$P_{2}$}
\put(150,-10){$P_{1}$}
\put(250,-10){$P_0$}
\put(350,-10){$P_{-1}$}
\put(450,-10){$P_{-2}$}

\put(220,-40){\makebox(0,0)[b]{\bf {Figure 5}}}
\end{picture}
\end{center}

\vspace{2cm}

As before, the parametrization is chosen to obey (\ref{onebraneagain})
so the curve becomes
\be
\sum_{n} q^{n^2/ 2} e^{\beta n z} P_0(x-nm)\,=\,0,
\label{zttwo}
\ee
with the leading term of $P_0(x)$ given by
\be
P_0(x)\,=\,\prod_{p=1}^{\infty} 
\left[(x+pm)(x-pm) \over (x+(p-\ha)m ) (x-(p-\ha)m) \right]^{2p}
 x \prod_{i=1}^N [x^2-a_i^2].
\label{ztone}
\ee
As in sec.~4.4, there are several decoupling limits 
that can be considered  to check eq.~(\ref{zttwo}).

\noindent
(i) This limit works exactly as in sec. 4.4,
to yield the curve for ${\cal N}=4$ SU$(N)$  + adjoint. 

\noindent
(ii) If we take 
\be
&& q \to 0\,\,\,(\tau \to i\infty)\,\,,\,\,\, \,\,\,\,
m \to  \infty\, \,,\,\,\,\,\Lambda^{b_0} =q m^{b_0}\,\,\,\,{\rm fixed}; \nn \\
&& b_0\,=\,4N_1-2N_2-2\,\,,\,\,\,\,\,\,\,N_1+N_2\,=\,N,
\label{decouplingtres}
\ee
the resulting curve agrees with the one for SO$(2N_1+1)$ + $N_2$ fundamentals.
(We disagree with the curve proposed in ref.~\cite{Yokono}, 
for reasons discussed in section 4.4.)

{}From (\ref{masterfcn}), we find
\be
S(x) \,=\, 
{\xf  x^2 (x+m)(x-m) \prod_{i=1}^N [(x-m)^2 -a_i^2] 
\prod_{i=1}^N [(x+m)^2 -a_i^2]
\over { (x+\ha m)^2 (x-\ha m)^2 \prod_{i=1}^N (x^2-a_i^2)^2}}.
\ee
The one-instanton contribution to the prepotential is given by
\be 
\xe \pi i {\cal F}_{\rm 1-inst} 
\, =\, \sum_{k=1}^{N} S_k(a_k)-2S_m(-\ha m),
\label{decouplingcinco} 
\ee 
with
\be
S_k(a_k)&=&
{\xf (a_k+m)(a_k-m) 
\prod_{i=1}^N [(a_k-m)^2 -a_i^2]
\prod_{i=1}^N [(a_k+m)^2 -a_i^2]
\over { 4 (a_k+\ha m)^2 (a_k-\ha m)^2 
\prod_{i\neq k}^N (a_k^2-a_i^2)^2}};\nn\\ \nn\\
S_m(-\ha m)&=&
{-3 m^2  \prod_{i=1}^N ({9 \over 4}  m^2 -a_i^2)
\over {  16 \prod_{i=1}^N ({1\over 4} m^2-a_i^2)}}.
\label{decouplingcuatro}
\ee
In the case of SO(5),
we will be able to test eq.~(\ref{decouplingcinco}) against
the result we will obtain in the next subsection
for Sp(4) with a massive adjoint hypermultiplet.

\eject

\noindent{\bf 4.6.} 
\underline{Sp$(2N)\,+\,{\rm adjoint}$}
\renewcommand{\theequation}{4.6.\arabic{equation}}
\setcounter{equation}{0}

Consider Sp$(2N)$ gauge theory 
with a matter hypermultiplet (mass $m$) in the adjoint representation.
The corresponding M-theory picture (Fig. 6) 
contains O$6^+$ planes on top of each NS 5-brane, 
O$6^-$ planes between each pair of NS 5-branes, 
and D4-branes. 
As in the previous section, 
there are additional fourbranes at mirror symmetric positions 
with respect to the O$6^-$ orientifolds 
that are not exhibited in Fig. 6 for clarity. 

\begin{center}
\begin{picture}(810,295)(10,10)


\put(10,192){\line(1,0){5}}
\put(20,192){\line(1,0){5}}
\put(30,192){\line(1,0){5}}
\put(40,192){\line(1,0){5}}
\put(50,192){\line(1,0){5}}
\put(60,192){\line(1,0){5}}
\put(70,192){\line(1,0){5}}
\put(80,192){\line(1,0){5}}
\put(90,192){\line(1,0){5}}
\put(100,192){\line(1,0){5}}
\put(10,180){$(\!x\!-\!2m\!-\!a_i\!)$}

\put(105,250){\line(1,0){2}}
\put(110,250){\line(1,0){5}}
\put(120,250){\line(1,0){5}}
\put(130,250){\line(1,0){5}}
\put(140,250){\line(1,0){5}}
\put(150,250){\line(1,0){5}}
\put(160,250){\line(1,0){5}}
\put(170,250){\line(1,0){5}}
\put(180,250){\line(1,0){5}}
\put(190,250){\line(1,0){5}}
\put(200,250){\line(1,0){5}}
\put(125,255){$(\!x\!-\!m\!+\!a_i\!)$}

\put(205,117){\line(1,0){2}}
\put(210,117){\line(1,0){5}}
\put(220,117){\line(1,0){5}}
\put(230,117){\line(1,0){5}}
\put(240,117){\line(1,0){5}}
\put(250,117){\line(1,0){5}}
\put(260,117){\line(1,0){5}}
\put(270,117){\line(1,0){5}}
\put(280,117){\line(1,0){5}}
\put(290,117){\line(1,0){5}}
\put(300,117){\line(1,0){5}}
\put(235,105){$(\!x\!-\!a_i\!)$}

\put(305,155){\line(1,0){2}}
\put(310,155){\line(1,0){5}}
\put(320,155){\line(1,0){5}}
\put(330,155){\line(1,0){5}}
\put(340,155){\line(1,0){5}}
\put(350,155){\line(1,0){5}}
\put(360,155){\line(1,0){5}}
\put(370,155){\line(1,0){5}}
\put(380,155){\line(1,0){5}}
\put(390,155){\line(1,0){5}}
\put(400,155){\line(1,0){5}}
\put(320,160){$(\!x\!+\!m\!+\!a_i\!)$}

\put(405,10){\line(1,0){2}}
\put(410,10){\line(1,0){5}}
\put(420,10){\line(1,0){5}}
\put(430,10){\line(1,0){5}}
\put(440,10){\line(1,0){5}}
\put(450,10){\line(1,0){5}}
\put(460,10){\line(1,0){5}}
\put(470,10){\line(1,0){5}}
\put(480,10){\line(1,0){5}}
\put(490,10){\line(1,0){5}}
\put(500,10){\line(1,0){5}}
\put(415,15){$(\!x\!+\!2m\!-\!a_i\!)$}


\put(105,5){\line(0,1){255}}
\put(205,5){\line(0,1){255}}
\put(305,5){\line(0,1){255}}
\put(405,5){\line(0,1){255}}
\put(103,271){$\NSone$}
\put(203,271){$\NStwo$}
\put(303,271){$\NSthree$}
\put(403,271){$\NSfour$}


\put(51,235){$\otimes$} 
\put(51,245){O$6^{-}$}
\put(10,235){$(\!x\!-\!2m\!)$}
\put(101,215){$\otimes$} 
\put(110,215){O$6^{+}$}
\put(201,175){$\otimes$}
\put(210,175){O$6^{+}$}
\put(301,135){$\otimes$}
\put(310,135){O$6^{+}$}
\put(401,95){$\otimes$}
\put(410,95){O$6^{+}$}
\put(62,215){$(\!x\!-\!\thh m\!)$}
\put(160,168){$(\!x\!-\!\ha m\!)$}
\put(260,130){$(\!x\!+\!\ha m\!)$}

\put(151,195){$\otimes$}
\put(151,205){O$6^{-}$}
\put(251,155){$\otimes$}
\put(251,165){O$6^{-}$}
\put(351,115){$\otimes$}
\put(351,125){O$6^{-}$}
\put(451,75){$\otimes$}
\put(451,85){O$6^{-}$}
\put(430,60){$(\!x\!+\!2m\!)$}

\put(130,184){$(\!x\!-\!m\!)$}
\put(362,85){$(\!x\!+\!\thh m\!)$}
\put(248,148){$\!x\!$}
\put(340,105){$(\!x+\!m\!)$}

\put(30,30){\vector(1,0){20}}
\put(30,30){\vector(0,1){20}}
\put(25,50){$x$}
\put(50,25){$t$}

\put(50,-10){$P_{2}$}
\put(150,-10){$P_{1}$}
\put(250,-10){$P_0$}
\put(350,-10){$P_{-1}$}
\put(450,-10){$P_{-2}$}

\put(220,-30){\makebox(0,0)[b]{\bf {Figure 6}}}

\end{picture}
\end{center}

\vspace{2cm}

Choosing the coefficient functions to obey (\ref{onebraneagain}), 
the curve for this theory becomes
\be
\sum_n  q^{n^2/2} e^{\beta n z} P_0(x-nm)\,=\,0,
\label{zseventeen}
\ee
where the leading term  of $ P_0(x)$ is
\be
P_0(x)\,=\,\prod_{p=1}^{\infty} 
\left[ (x+(p-\ha)m ) (x-(p-\ha)m) \over (x+pm)(x-pm) \right]^{2p}
\prod_{i=1}^N [x^2-a_i^2].
\label{zfifteen}
\ee
Notice that the non-dynamical factors of (\ref{zfifteen}) are the inverse
of those in (\ref{ztwelve}).  
The curve (\ref{zseventeen}) is manifestly invariant under
$z \to z+2\omega_2$ and $x \to x+m$.

As in section 4.4, we can consider some decoupling limits 
to check eq.  (\ref{zseventeen}). 

\noindent
(i) This limit is again the same as in sec.~4.4.

\noindent
(ii) If we consider 
\be
&& q \to 0\,\,\,(\tau \to i\infty)\,\,,\,\,\, \,\,\,\,
m \to \infty \,\, ,\,\,\,\,\Lambda^{b_0} =q m^{b_0}\,\,\,\,{\rm fixed}; \nn \\
&& b_0\,=\,4N_1-2N_2+4\,\,,\,\,\,\,\,\,\,N_1+N_2\,=\,N,
\label{decouplingdos}
\ee
the curve we obtain agrees with the one 
corresponding to ${\cal N}=2$ Sp$(2N_1)$ + $N_2$ fundamentals, as it should.
(We disagree with the curve proposed in ref.~\cite{Yokono}, 
for the reasons given in section 4.4.)

{}From (\ref{masterfcn}), we obtain
\be 
S(x) = {\xf  (x+\ha m)^2 (x-\ha m)^2 \prod_{i=1}^N [(x-m)^2 -a_i^2]
\prod_{i=1}^N [(x+m)^2 -a_i^2]
\over { x^4\prod_{i=1}^N (x^2-a_i^2)^2}}.
\label{spnprepot}
\ee
Using the methods of sec.~2 of this paper,
one obtains the following one-instanton contribution to the prepotential 
\be
\xe \pi i {\cal F}_{\rm 1-inst}\,=\,-2 [\quarS_0(0)]^{1/2},
\label{zeighteen}
\ee
where
\be
\quarS_0(0)\,=\,
{\xf   (-m^2)^2 \prod_{i=1}^N (m^2-a_i^2)^2 \over 16 \prod_{i=1}^N (-a_i{}^2)^2}.
\label{porqueno}
\ee

We can check eqs. (\ref{zeighteen}) and (\ref{porqueno}) 
by specializing to Sp(2) and comparing with SU(2) 
plus adjoint hypermultiplet with mass $m$. 
The corresponding prepotentials agree up to a rescaling 
and a moduli-independent additive constant.
We can also test eqs. (\ref{zeighteen}) and (\ref{porqueno}) 
for Sp(4) with a massive adjoint hypermultiplet
against eqs. (\ref{decouplingcinco}) and (\ref{decouplingcuatro}) 
for SO(5) with a massive adjoint hypermultiplet. 
Using the change of variables (\ref{thanksgiving})
relating the moduli of SO(5) and Sp(4), 
one can show that the two results agree up to a rescaling,
and a moduli-independent additive constant. 
This is actually a consistency check of our methods rather 
than a truly independent test.

\eject

\noindent{\bf 4.7.} 
\underline{Sp$(2N)$ + 1 antisymmetric + $N_f$ fundamentals} 
\renewcommand{\theequation}{4.7.\arabic{equation}}
\setcounter{equation}{0}

Consider Sp$(2N)$ gauge theory 
with a matter hypermultiplet (mass $m$) in the antisymmetric representation,
and $N_f \leq 4$ matter hypermultiplets (masses $M_j$) in the fundamental
representation.
The M-theory picture in Fig. 7
contains O$6^-$ planes on top of each NS 5-brane, 
O$6^-$ planes between each pair of NS 5-branes, 
together with D4-branes and D6-branes.
As in previous examples, 
there are additional D4-branes and D6-branes
at mirror symmetric positions with respect to the O$6^{-}$
orientifolds that are not depicted in Fig. 7.

\begin{center}
\begin{picture}(810,295)(10,10)


\put(10,192){\line(1,0){5}}
\put(20,192){\line(1,0){5}}
\put(30,192){\line(1,0){5}}
\put(40,192){\line(1,0){5}}
\put(50,192){\line(1,0){5}}
\put(60,192){\line(1,0){5}}
\put(70,192){\line(1,0){5}}
\put(80,192){\line(1,0){5}}
\put(90,192){\line(1,0){5}}
\put(100,192){\line(1,0){5}}
\put(10,180){$(\!x\!-\!2m\!-\!a_i\!)$}

\put(105,250){\line(1,0){2}}
\put(110,250){\line(1,0){5}}
\put(120,250){\line(1,0){5}}
\put(130,250){\line(1,0){5}}
\put(140,250){\line(1,0){5}}
\put(150,250){\line(1,0){5}}
\put(160,250){\line(1,0){5}}
\put(170,250){\line(1,0){5}}
\put(180,250){\line(1,0){5}}
\put(190,250){\line(1,0){5}}
\put(200,250){\line(1,0){5}}
\put(125,255){$(\!x\!-\!m\!+\!a_i\!)$}

\put(205,117){\line(1,0){2}}
\put(210,117){\line(1,0){5}}
\put(220,117){\line(1,0){5}}
\put(230,117){\line(1,0){5}}
\put(240,117){\line(1,0){5}}
\put(250,117){\line(1,0){5}}
\put(260,117){\line(1,0){5}}
\put(270,117){\line(1,0){5}}
\put(280,117){\line(1,0){5}}
\put(290,117){\line(1,0){5}}
\put(300,117){\line(1,0){5}}
\put(235,105){$(\!x\!-\!a_i\!)$}

\put(305,155){\line(1,0){2}}
\put(310,155){\line(1,0){5}}
\put(320,155){\line(1,0){5}}
\put(330,155){\line(1,0){5}}
\put(340,155){\line(1,0){5}}
\put(350,155){\line(1,0){5}}
\put(360,155){\line(1,0){5}}
\put(370,155){\line(1,0){5}}
\put(380,155){\line(1,0){5}}
\put(390,155){\line(1,0){5}}
\put(400,155){\line(1,0){5}}
\put(320,160){$(\!x\!+\!m\!+\!a_i\!)$}

\put(405,10){\line(1,0){2}}
\put(410,10){\line(1,0){5}}
\put(420,10){\line(1,0){5}}
\put(430,10){\line(1,0){5}}
\put(440,10){\line(1,0){5}}
\put(450,10){\line(1,0){5}}
\put(460,10){\line(1,0){5}}
\put(470,10){\line(1,0){5}}
\put(480,10){\line(1,0){5}}
\put(490,10){\line(1,0){5}}
\put(500,10){\line(1,0){5}}
\put(415,15){$(\!x\!+\!2m\!-\!a_i\!)$}


\put(105,5){\line(0,1){255}}
\put(205,5){\line(0,1){255}}
\put(305,5){\line(0,1){255}}
\put(405,5){\line(0,1){255}}
\put(103,271){$\NSone$}
\put(203,271){$\NStwo$}
\put(303,271){$\NSthree$}
\put(403,271){$\NSfour$}


\put(51,235){$\otimes$} 
\put(51,245){O$6^{-}$}
\put(10,235){$(\!x\!-\!2m\!)$}
\put(101,215){$\otimes$} 
\put(110,215){O$6^{-}$}
\put(201,175){$\otimes$}
\put(210,175){O$6^{-}$}
\put(301,135){$\otimes$}
\put(310,135){O$6^{-}$}
\put(401,95){$\otimes$}
\put(410,95){O$6^{-}$}
\put(62,215){$(\!x\!-\!\thh m\!)$}
\put(160,168){$(\!x\!-\!\ha m\!)$}
\put(260,130){$(\!x\!+\!\ha m\!)$}

\put(151,195){$\otimes$}
\put(151,205){O$6^{-}$}
\put(251,155){$\otimes$}
\put(251,165){O$6^{-}$}
\put(351,115){$\otimes$}
\put(351,125){O$6^{-}$}
\put(451,75){$\otimes$}
\put(451,85){O$6^{-}$}
\put(430,60){$(\!x\!+\!2m\!)$}

\put(151,100){\framebox(5,5){$\cdot$}}
\put(130,90){$(\!x\!-\!M_j\!-\!m\!)$}
\put(251,250){\framebox(5,5){$\cdot$}}
\put(235,238){$(\!x\!+\!M_j\!)$}
\put(351,20){\framebox(5,5){$\cdot$}}
\put(331,35){$(\!x\!-\!M_j\!+\!m\!)$}

\put(130,184){$(\!x\!-\!m\!)$}
\put(362,85){$(\!x\!+\!\thh m\!)$}
\put(248,148){$\!x\!$}
\put(340,105){$(\!x+\!m\!)$}

\put(30,30){\vector(1,0){20}}
\put(30,30){\vector(0,1){20}}
\put(25,50){$x$}
\put(50,25){$t$}

\put(50,-10){$P_{2}$}
\put(150,-10){$P_{1}$}
\put(250,-10){$P_0$}
\put(350,-10){$P_{-1}$}
\put(450,-10){$P_{-2}$}

\put(220,-40){\makebox(0,0)[b]{\bf {Figure 7}}}
\end{picture}
\end{center}

\vspace{4cm}

There is one NS 5-brane per unit cell ($\arh=2$), 
so we choose a parametrization for the coefficient functions
obeying (\ref{onebrane}), yielding the curve
\be
\sum_{n=-\infty}^\infty L^{2n^2}\,t^n\,  J_0(x-nm)\, P_0(x-nm)\, =0, 
\label{xxxuno}
\ee
where $L^2=\Lambda^{4-N_f}$,
with the $D6$ branes responsible for the function
\be
J_{0}(x)=
\prod_{j=1}^{N_f}
\prod_{p=1}^{\infty} \,\biggl[ (x - p m - M_j) (x - p m + M_j)
                              (x + p m - M_j) (x + p m + M_j) \biggr]^{p/2},
\label{ztwenty}
\ee
and the leading term (in $L$) of $P_0(x)$ given by
\be
P_0(x)\,=\,\prod_{p=1}^{\infty} 
\left[ (x+(p-\ha)m ) (x-(p-\ha)m) (x+pm) (x-pm) \right]^{-2p}
\prod_{i=1}^N [x^2-a_i^2].
\label{znineteen}
\ee
One may verify that in the $m \to \infty$ limit, 
the curve reduces to that for Sp$(2N)$ with $N_f$ fundamental hypermultiplets.

Using eq.~(\ref{masterfcn}), one obtains
\be
S(x) = 
 {\xf  \prod_{i=1}^N [(x-m)^2 -a_i^2] \prod_{i=1}^N [(x+m)^2 -a_i^2]
\prod_{j=1}^{N_f}(x^2-M_j^2)
\over { x^4 (x+\ha m)^2 (x-\ha m)^2  \prod_{i=1}^N (x^2-a_i^2)^2}}.
\ee
Using the methods of section 2,
one obtains the one-instanton contribution to the prepotential 
is given by 
\be
 \xe \pi i {\cal F}_{\rm 1-inst}\,= \,-2 [\quarS_0(0)]^{1/2},
 \label{xxxxdos}
 \ee
 where
 \be
 \quarS_0(0)\,=\,{\xf 16 \prod_{i=1}^N (m^2-a_i^2)^2 \prod_{j=1}^{N_f}(-M_j^2) 
 \over (-m^2)^2 \prod_{i=1}^N (-a_i{}^2)^2}.
 \label{xxxtres}
 \ee

Several checks may be made of this result.
For Sp(2), eqs.~(\ref{xxxxdos}) and (\ref{xxxtres}) yield
\be
\xe \pi i {\cal F}_{\rm 1-inst}\,
=\, 8 i^{N_f}{\prod_{j=1}^{N_f} M_j\over m^2}
-\, 8 i^{N_f}{\prod_{j=1}^{N_f} M_j\over a_1{}^2}.
\label{xvi}
\ee
which agrees with the one-instanton prepotential
for Sp(2) with $N_f$ fundamental hypermultiplets
(up to a moduli-independent additive constant and an overall rescaling),
as expected, since the antisymmetric representation of Sp(2) is trivial.
The result for Sp(4) with 1 antisymmetric hypermultiplet
and no fundamental hypermultiplets
agrees with the one-instanton prepotential
for SO(5) with one fundamental hypermultiplet \cite{DHokerKricheverPhong3} 
(see Table 1) up to a rescaling and a moduli independent
additive constant, after making the change of variables~(\ref{thanksgiving}).

\vskip.3in
\noindent{\bf 5. ~Comparison to elliptic curves with zero global mass}
\renewcommand{\theequation}{5.\arabic{equation}}
\setcounter{equation}{0}

Complementary to our strategy of deriving curves from the
M-theory pictures of Uranga \cite{Witten}--\cite{Uranga},
there exist methods developed for elliptic models by Donagi and 
Witten \cite{DonagiWitten}, Uranga \cite{Uranga},
Gukov and Kapustin \cite{GukovKapustin}, and others. 
Although a curve results from their considerations, 
the extraction of the instanton expansion has not been carried out 
for these curves.
In short, the issue is how to extract $P_n(x)$ and $J_n(x)$,
and from these, $S(x)$, 
from the curves of the Donagi-Witten type. 
In this section, we accomplish this for two non-trivial 
models with zero beta function and zero global mass. 

The curves for theories in sec.~4 
with only one NS 5-brane per unit cell become trivial 
({\ie}, factorize into a function of $x$ and a function of $t$ \cite{Uranga})
when the global mass $m$ vanishes.
This yields $S(x)=$ constant, 
and a vanishing one-loop and instanton prepotential.
On the other hand,
the curves for theories with two NS 5-branes per unit cell,
{\it viz.}, SU$(N)$ with two antisymmetric hypermultiplets
and four fundamental hypermultiplets,
and SU$(N)$ with one antisymmetric and one symmetric hypermultiplet,
both have the non-trivial limit (\ref{specialcase})
when the global mass $\Delta$ vanishes.
We show that the curves obtained by Gukov and Kapustin, 
and Uranga, respectively, for these two models agree, 
after a suitable change of variables, with eq. (\ref{specialcase}).

\vskip.3in
\noindent{\bf 5.1.}
\underline{SU$(N)+\,2\,$ antisymmetric $\,+\,4$ fundamentals}
\renewcommand{\theequation}{5.1.\arabic{equation}}
\setcounter{equation}{0}

Consider 
the SU$(N)$ gauge theory,
with two antisymmetric hypermultiplets (masses $m_1$ and $m_2$)
and four fundamental hypermultiplets (masses $M_j$),
with their masses related by $m_1=m_2=2M_j$.
This is an elliptic model with zero global mass. 
Gukov-Kapustin \cite{GukovKapustin} give the curve for  SU$(2n)$
\be
v^{2n}\,+\,f_1(x,y) v^{2n-1}\,+ \,\cdots\, +\, f_{2n}(x,y) \,=\,0,
\label{htfour}
\ee
with the coefficient functions 
\be
f_{2j}(x,y)&=&A_j,\nn\\
f_{2j-1}(x,y)&=&{{yB_j}\over (x-e_3)}\,=\,{{(x-e_1)(x-e_2)}\over y} B_j ,
\label{htfive}
\ee
where $v = x_4 + i x_5$ is what we called $x$ in earlier sections,
$A_j$ and $B_j$ are constants, 
and $x$ and $y$ parametrize the torus base space via
\be
y^2\,=\, (x-e_1)(x-e_2)(x-e_3).
\label{xuno}
\ee
We will assume that precisely the same functions (\ref{htfive}) appear in the 
curve for SU$(N)$ with  $N$ odd, 
so the curve for any $N$ takes the form:
\be
v^{N}\,+\,f_1(x,y) v^{N-1}\,+ \,\cdots\, +\, f_{N-1}(x,y) v\,+\, 
f_{N}(x,y) \,=\,0.
\label{htfour2}
\ee 
Let 
\be
H_0(v)  &=& \prod_{j=1}^{N}(v-a_j-\nha m) = \sum_{j=0}^N u_j v^{N-j},\nn\\
H_1(v)  &=& H_0(-v)= (-1)^N \prod_{j=1}^{N}(v+a_j+\nha m) = (-1)^N 
\sum_{j=0}^N (-1)^j u_j v^{N-j},~~~~~~~~~~~~~~~~~~
\label{hteight}
\ee
which defines the $u_j$ as gauge invariant combinations 
of $m$ and the order parameters $a_j$ (with $u_0=1$). 
Let also 
\be
H_{\rm even} &=& \sum_{i~{\rm even}}u_{i}v^{N-i},\qquad\qquad\qquad\,\, 
H_{\rm odd} \,= \,\sum_{i~{\rm odd}}u_{i}v^{N-i},\nn \\ 
H_0 &=& H_{\rm even}+H_{\rm odd},\qquad\qquad 
(-1)^N H_1 \,=\, H_{\rm even}-H_{\rm odd}.
\label{hteighti}
\ee
Then  (\ref{htfour2}) can 
be written as 
\be
H_{\rm even}(v)+{(x-e_1)(x-e_2)\over y}H_{\rm odd}(v)=0, 
\label{xdos}
\ee
where we identify $u_{2j}=A_j$, and $u_{2j-1}=B_j$.

When $\sum_{i}e_i=0$ in eq. (\ref{xuno}), 
the variables $x$ and $y$ are related to the Weierstrass elliptic functions by 
\be
x\,=\,{\wp}(z)\qquad 2y\,=\, {\wp}'(z),
\label{nttwo}
\ee
where $z$ parametrizes the base torus
with the identifications $z \equiv z + 2 \omega_1 \equiv z + 2 \omega_2$,
and therefore 
\be
{y\over (x-e_1)(x-e_2)}\,=\,-ic\,{\theta_1(\nu\vert\tau)
\theta_4(\nu\vert\tau)\over
\theta_2(\nu\vert\tau)\theta_3(\nu\vert\tau)}\,=\, 
-i c{\theta_1(\nu\vert{\tau\over2})\over
\theta_2(\nu\vert{\tau\over2})}=c f(\nu|\tau),
\label{xtres}
\ee
where  $\nu = z/2 \omega_1$ and
\be
c=-{2i\omega_1\over\pi\theta_4(0|\tau)^2},
\label{c}
\ee 
\be
f(\nu|\tau)=-i{\theta_1(\nu\vert{\tau\over2})\over
\theta_2(\nu\vert{\tau\over2})},
\label{f}
\ee
and \cite{Polchinski}  
\be
\theta_1(\nu|\tau) &=& i \sum_{n=-\infty}^{\infty} (-1)^n 
e^{i \pi \tau (n - \tshalf)^2} e^{2 \pi i \nu (n - \tshalf)}, \nn\\ 
\theta_2(\nu|\tau) &=&  \sum_{n=-\infty}^{\infty} 
e^{i \pi \tau (n - \tshalf)^2} e^{2 \pi i \nu (n - \tshalf)}, \nn\\ 
\theta_3(\nu|\tau) &=&  \sum_{n=-\infty}^{\infty} 
e^{i \pi \tau n^2} e^{2 \pi i \nu n}, \nn\\ 
\theta_4(\nu|\tau) &=&  \sum_{n=-\infty}^{\infty} (-1)^n 
e^{i \pi \tau n^2} e^{2 \pi i \nu n}.
\label{thetafcns}
\ee 
In view of eqs. (\ref{xtres})-(\ref{f}) we can rewrite the curve (\ref{xdos}) 
as follows
\be
H_{\rm even}(v)+[c f(\nu|\tau)]^{-1}H_{\rm odd}(v)=0. 
\label{xdosi}
\ee
One may redefine the gauge invariant moduli as follows:
\be
u'_j=\cases{u_j,\qquad \,\,\,\,\,\, j~ {\rm even},\cr 
u_j/c,\qquad j~ {\rm odd},}
\label{xnueve}
\ee
which is valid since (\ref{c}) is independent of $\nu$ and $ v$.
Next, we shift $\nu\to \nu +{\tau\over4}+{1\over2}$ if $N$ is even, and 
$\nu\to \nu +{\tau\over4}$ if $N$ is odd. 
Since
\be
f(\nu+{\tau\over 4}|\tau)={\theta_4(\nu\vert{\tau\over2})\over
\theta_3(\nu\vert{\tau\over2})}, \qquad 
f(\nu+{\tau\over 4}+{1\over2}|\tau)={\theta_3(\nu\vert{\tau\over2})\over
\theta_4(\nu\vert{\tau\over2})},
\label{fi}
\ee
this converts the curve (\ref{xdosi}) into
\be
&&\theta_3(\nu\vert{\tau\over2})\,H_{\rm even}(v)+
\theta_4(\nu\vert{\tau\over2})
\,H_{\rm odd}(v)=0,\qquad N~{\rm even},\nn \\
&&\theta_4(\nu\vert{\tau\over2})\,H_{\rm even}(v)+
\theta_3(\nu\vert{\tau\over2})
\,H_{\rm odd}(v)=0,\qquad N~{\rm odd}.
\label{xdosii}
\ee
Using the identities
\be
\theta_3(\nu\vert{\tau\over2})&=&\theta_3(2\nu\vert 2\tau)+
\theta_2(2\nu\vert 2\tau),
\nn\\
\theta_4(\nu\vert{\tau\over2})&=&\theta_3(2\nu\vert 2\tau)-
\theta_2(2\nu\vert 2\tau),
\label{xcinco}
\ee
eq. (\ref{xdosii}) becomes, for both even or odd $N$,
\be
H_0(v)\,\theta_3(2 \nu|2 \tau)\,+\,
H_1(v)\,\theta_2(2 \nu|2 \tau)\,=\,0.
\label{htnine}
\ee
This result exactly agrees with our result (\ref{specialcase}), 
when we set $v=x+\nha m$. 
This supports the validity of the methods used in
obtaining the curves in sec.~4.
Further, we know that for this case of zero global mass, 
the functions $P_n(x)$ do {\it not} have any subleading 
terms, from two points of view: 
(a) the exact agreement of the leading terms with eq.~(\ref{htnine}), 
and 
(b) the absence of ``subtractions" for the one-instanton prepotential. 
\eject

\vskip.3in
\noindent{\bf 5.2.} 
\underline{SU$(N)\,+\,1\, {\rm antisymmetric}\,+\,1\, {\rm symmetric}$}
\renewcommand{\theequation}{5.2.\arabic{equation}}
\setcounter{equation}{0}

For zero global mass, Uranga \cite{Uranga} gives the curve (\ref{htfour2})  
with 
\be
f_{2j}(x,y)&=&C_j,\nn\\
f_{2j-1}(x,y)&=&{{yD_j}\over (x-e_1)(x-e_2)}\,=\,{{(x-e_3)}\over y} D_j. 
\label{xdiez}
\ee
The curve for this case, with $m_1=m_2$, is therefore 
\be
 H_{\rm even}(v)+ {y\over (x-e_1)(x-e_2)} H_{\rm odd}(v)=0,
\label{xonce}
\ee
where we identify $u_{2j}=C_j$ and  $u_{2j-1}=D_j$. 
Following (\ref{nttwo})-(\ref{xdosi}), we have  
\be
H_{\rm even}(v)+ cf(\nu|\tau) H_{\rm odd}(v)=0,
\label{xdoce}
\ee
with $c$ and $f(\nu|\tau)$ defined by eqs. (\ref{c}) 
and (\ref{f}) respectively. 
We redefine the gauge invariant moduli as 
\be
u'_j=\cases{u_j,\qquad \,\,j~ {\rm even},\cr 
c\,u_j,\qquad j~ {\rm odd}.} 
\label{xtrece}
\ee 
Further, we shift $\nu\to \nu +{\tau\over4}$ if $N$ is even, and 
$\nu\to \nu +{\tau\over4}+{1\over2}$ if $N$ is odd.
(Note that the shifts for even and odd $N$ are the reverse
of those in the previous section.)
This transforms the curve (\ref{xdoce}) into eq.~(\ref{htnine}),
agreeing with the result (\ref{specialcase}) obtained in sec.~4, 
with no subleading terms for the $P_n(x)$.

Thus, for  SU($N$) $+$ 2~anti. $+$ 4 fund. 
and SU($N$) $+$ 1 sym. $+$ 1 anti. 
with zero global mass in both cases 
(and $m=2M_j$ for the former),
the curves are identical, 
{\ie}, the two theories have identical prepotentials. 
This was already noted in Table 4, at the one-instanton 
level. 
One may wonder from the \hbox{M-theory} point of view why this 
has occurred. 
{}From Fig.~1 for SU($N$) + 2~anti. + 4 fund., 
we see that, if all the masses are equal,
the positions of the D6-branes 
have the same value of $v$ ($x$ in the figure) as that 
of the two O$6^{-}$ orientifold planes. 
One may then bring the four D6-branes (plus mirrors)  
coincident with one 
of the O$6^{-}$ planes, converting this 
effectively to an O$6^{+}$ plane. The resulting configuration is 
that of SU($N$) $+$ 1~sym. $+$ 1~anti. 
for zero global mass. Hence, the identity of the low-energy theories 
with zero global mass could have 
been anticipated. 
On the other hand, if there are global masses,
this construction is not possible and the 
curves no longer coincide. 

\ieject
 
\vskip.3in
\noindent{\bf 6. ~Comparison with the curve of Donagi and Witten}
\renewcommand{\theequation}{6.\arabic{equation}}
\setcounter{equation}{0}

The M-theory picture corresponding to the SU$(N)$ gauge theory
with massive adjoint hypermultiplet \cite{Witten}
is described in sec. 4.3 (see Figs. 3.a and 3.b).
Using this, we obtained the curve (\ref{zeleven})
\be
\sum_{n=-\infty}^{\infty}\,(-1)^n q^{n(n-1)/2}\,e^{\beta nz}\, H(v-nm)=0,
\label{xi}
\ee
where 
\be
H(v)=\prod_{i=1}^{N}\, (v-a_i),
\label{xii}
\ee
with $v=x_4+ix_5$  (previously referred to as $x$), and  $q=e^{2\pi i \tau}$.
As before, $z$ parametrizes the base torus with the identifications 
$z \equiv z + 2 \omega_1 \equiv z + 2 \omega_2$,
and throughout this section we fix $\omega_1=-\pi i$ (hence $\beta=1$)
for convenience. 
This is exactly the curve derived 
in ref. \cite{DHokerPhong,MartinecWarner},
for the Calogero-Moser model.

On the other hand,
Witten \cite{Witten} shows that the curve for this model is 
precisely that derived by Donagi and Witten \cite{DonagiWitten}
in the context of the integrable Hitchin system
\be
F(v,x,y)\,=\,\sum_{j=0}^N A_j P_{N-j}(v),
\label{hsixteen}
\ee
where $A_j$ are 
gauge invariant polynomials in $a_i$ and $m$, 
and where $x$ and $y$ are related by (\ref{xuno}).
They show that 
\be
P_n(v)\,=\,\sum_{i=0}^n\,
\left( n \atop i \right) f_i \,\, v^{n-i}.
\label{heighteen}
\ee
It can be shown that 
\be
F(v,x,y)\,=\,\sum_{j=0}^N {m^j \over j!}\,\, f_j\,\, H^{(j)}(v),
\label{hnineteen}
\ee
where
\be
H^{(j)}(v)= {d^j H(v) \over dv^j}.
\label{htwenty}
\ee
Explicit calculation gives, using $\sum_i e_i=0$ in (\ref{xuno})
\be
f_0&=&1,\,\,\,\,\,\,\,\,\,\,\,\,\,\,\,\,\qquad
f_1\,=\,0,\,\,\,\,\,\,\,\,\,\,\,\,\,\qquad
f_2\,=\,-x,\,\,\,\,\,\,\,\,\nn\\
f_3&=&2y,\,\,\,\,\,\,\,\,\,\,\,\,\,\qquad
f_4\,=\,-3x^2,\,\,\,\,\,\,\,\,\,\,\,\ f_5=4xy,\,\,\,\qquad {\rm etc}.
\label{htone}
\ee

We will establish the connection between the curves 
(\ref{xi}) and (\ref{hnineteen}),
which has not been done explicitly previously.
The curve (\ref{xi}) can be recast as 
\be
\sum_{j=0}^{N}\,{(-m)^j \over j!}\,h_j(z)  H^{(j)} (v-{\tshalf}m)\,=\,0, 
\label{hthirteen}
\ee 
where
\be
h_j(z)={1 \over {\theta_1({z \over -2 \pi i}|\tau)}}{\pr^j \over {\pr z^j}}
\,\theta_1({z \over - 2 \pi i}|\tau), 
\label{hfourteen}
\ee 
with $\theta_1$ defined in eq. (\ref{thetafcns}).
Making the change of variables 
\be 
v\longrightarrow v + m h_1(z) + {\tshalf} m,
\label{mcbeal}
\ee
eq. (\ref{hthirteen}) becomes
\be
\sum_{j=0}^N {m^j \over j!} \, \, \tf_j(z) \, \,H^{(j)}(v)\,=\, 0,
\label{jaito}
\ee
with
\be
\tf_j (z) = \sum_{i=0}^j \left(j \atop i\right) (-1)^i h_i(z)
h_1(z)^{j-i}.
\label{jaitito}
\ee
Explicitly
\be
\tf_0&=&h_0\,=\,1,\nn\\
\tf_1&=&0,\nn\\
\tf_2&=&h_2-h_1{}^2   \,=\, h_1^\prime 
\,=\, - \wp(z) -{\pi^2\over 12\omega_1{}^2} E_2
\,=\, - \wp(z) +{\textstyle{1\over 12}} E_2 ,\nn\\
\tf_3&=&-h_3+3h_2h_1-2h_1{}^3= - h_1^{\prime\prime} = 2y,\nn\\
\tf_4&=&h_4-4h_3h_1+6h_2h_1{}^2-3h_1{}^4\,
=\,-3x^2-{\ha}\,E_2\,x
+{\textstyle{1\over48 }}E_2{}^2
+{\ha}\, g_2,\nn\\
\tf_5&=&-h_5+5h_4h_1-10h_3h_1{}^2+10h_2h_1{}^3-4h_1{}^5\,=\,
4xy+{\textstyle{5\over 3}}\,E_2\,y,~~~{\rm etc.}
\label{htthree}
\ee 
where $E_2$ is the Eisenstein  series of weight two, 
and $g_2$ is defined via the Weierstrass form 
\be
y^2=x^3-{\textstyle {1\over 4}}g_2 x- {\textstyle {1\over 4}} g_3,
\ee
of the elliptic curve (\ref{xuno}).
The relations (\ref{nttwo})  have been used in computing (\ref{htthree}). 

Comparing (\ref{htthree}) with (\ref{htone}),
we find that $f_j$ and $\tf_j$ differ by the
$\tau$-dependent change of basis
\be
\tf_0&=&f_0,  \nn\\
\tf_1&=&f_1,  \nn\\
\tf_2&=&f_2 + {\textstyle {1\over 12}} E_2(\tau) f_0,\nn\\
\tf_3&=&f_3,	\nn\\
\tf_4&=&f_4 + {\textstyle{1\over 2}}\,E_2(\tau) ~f_2
+\left[{\textstyle{1\over48}}E_2(\tau)^2+{\ha}g_2\right]~f_0,\nn\\
\tf_5&=&f_5 + {\textstyle{5\over6}} E_2(\tau) ~f_3,\qquad\qquad {\rm etc.}
\ee
This is similar but not identical to the comparison of the
spectral curve with the Donagi-Witten curve
made by Itoyama and Morozov \cite{ItoyamaMorozov},
to which we refer the reader for further discussion of this issue.

In this section and in section 5 
we have dealt with elliptic models that have no subleading terms 
for the coefficient functions $P_n(x)$. 
It remains an open question  how to carry out the analogous studies 
for elliptic models with a global mass and 
non-vanishing subleading terms.

\vskip.3in
\noindent{\bf 7. ~Concluding Remarks}
\renewcommand{\theequation}{7.\arabic{equation}}
\setcounter{equation}{0}

In this paper, we have provided a rather comprehensive presentation
of the relationship of elliptic models to M-theory, and related topics. 
There are, however, a number of topics which deserve further attention,
as they represent issues not completely understood. 

First, the information in Tables~1~and~3 would appear
to have an underlying group-theoretic explanation.   That is, 
given the group and matter content, one should be able to 
predict the factors in Table~3, without appealing to a SW curve. 
We know of no such explanation.

Second, for a number of elliptic models, only the leading
terms of the coefficient functions $P_n(x)$ are known.
(The leading term is that with the lowest power of $\Lambda$ or $q$ 
for a given power of $t$.)
This occurs for all models in section 4 
with non-zero global mass except SU$(N)$ + adjoint.
In these models,  the subleading terms are not known. 
Some non-elliptic models,
namely SU$(N)$ + 1 anti. + $N_f$ fund.
and Sp$(2N)$ + $N_f$ fund. (treated in section~2),
also contain subleading terms, which are explicitly known in these cases.
Comparison with Table~2 correlates models with subleading terms
with entries in the table in which ${\cal F}_{\rm 1-inst}$ 
involves terms other than $S_k(a_k)$
({\it e.g.},   $[\quarS_0(0)]^{1/2}$, 
or subtractions of factors such as $S_m(-\ha m)$).
These additional terms are closely connected to 
the residue functions $R_k(x)$, {\it e.g.,} eq.~(\ref{dosocho}), 
which originate in the subleading terms,
and which generate the additional terms via identities such as 
eq.~(\ref{doscatorce})
(leading in the case of Sp$(2N)$ + $N_f$ fund. to eq.~(\ref{dosdiezsiete})). 
Thus, the existence of subleading terms 
and the need for the subtractions 
listed in Table~2 are different aspects of the same issue.
The subtractions listed in Table~2 for models described in
sections 4.1, 4.2,  and 4.4 through 4.7, 
do not, however, give enough information to determine the subleading terms.
A greater understanding of these subleading terms would be desirable.

Finally, there are a number of elliptic models described in section~4
which have not been identified with known integrable models. It would
improve our understanding of the subject  if these connections
could be made. 

In conclusion, this paper represents considerable progress
toward a unified description of elliptic models associated with M-theory.
Nevertheless, as outlined above, there remain a number of interesting
issues to consider. 
\vspace{.15in}

\begin{center}
{\bf ~Acknowledgements} 
\end{center}

We wish to thank Henric Rhedin for his collaboration in the 
initial stages of this study, and his participation in 
the earlier work of this series.
S.G.N. would like to thank 
A. Giveon, A. Hanany, K. Hori, and D. Kutasov
for helpful discussions.

\end{document}